\documentclass[apj]{emulateapj}

\usepackage{color}

\usepackage{lipsum}

\newcommand{\msun}{\,\ifmmode {\rm M}_\odot \else ${\rm M}_\odot$\fi}
\newcommand{\softening}{\ifmmode \epsilon_{\rm soft} \else $\epsilon_{\rm soft}$\fi}
\newcommand{\Doff}{\ifmmode D_{\rm off} \else $D_{\rm off}$\fi}
\newcommand{\costheta}{\ifmmode |\!\cos \, \theta| \else $|\!\cos \, \theta|$\fi}

\lefthead{Kuhlen et al.}
\righthead{}

\begin{document}

\title{An Off-center Density Peak in the Milky Way's Dark Matter Halo?}

\author{Michael Kuhlen\altaffilmark{1}, Javiera Guedes\altaffilmark{2}, Annalisa Pillepich\altaffilmark{3}, Piero Madau\altaffilmark{3}, and Lucio Mayer\altaffilmark{4}}

\affil{$^1$Theoretical Astrophysics Center, University of California Berkeley, Hearst Field Annex, Berkeley, CA 94720 \\
       $^2$ETH Zurich, Institute for Astronomy, Wolfgang-Pauli-Strasse 27, Zurich 8049, Switzerland \\
       $^3$Department of Astronomy \& Astrophysics, University of California Santa Cruz, 1156 High St., Santa Cruz, CA 95064 \\
       $^4$University of Zurich, Institute for Theoretical Physics, Zurich 8057, Switzerland}

\email{mqk@astro.berkeley.edu}

\begin{abstract}
We show that the position of the central dark matter density peak may be expected to differ from the dynamical center of the Galaxy by several hundred parsec. In Eris, a high resolution cosmological hydrodynamics simulation of a realistic Milky-Way-analog disk galaxy, this offset is \mbox{300 - 400} pc ($\sim 3$ gravitational softening lengths) after $z=1$. In its dissipationless dark-matter-only twin simulation ErisDark, as well as in the Via Lactea II and GHalo simulations, the offset remains below one softening length for most of its evolution. The growth of the DM offset coincides with a flattening of the central DM density profile in Eris inwards of $\sim 1$ kpc, and the direction from the dynamical center to the point of maximum DM density is correlated with the orientation of the stellar bar, suggesting a bar-halo interaction as a possible explanation. A dark matter density offset of several hundred parsec greatly affects expectations of the dark matter annihilation signals from the Galactic Center. It may also support a dark matter annihilation interpretation of recent reports by \citet{weniger_tentative_2012} and \citet{su_strong_2012} of highly significant 130 GeV gamma-ray line emission from a region $1.5^\circ$ ($\sim200$ parsec projected) away from \mbox{Sgr A*} in the Galactic plane.
\end{abstract}


\section{Introduction}

Dissipationless (dark matter only) N-body simulations predict a nearly universal density profile of dark matter (DM) halos, the so-called NFW profile \citep{navarro_universal_1997}, which features a central $1/r$ density cusp. If the cooling and condensation of gas inside these halos \citep{white_core_1978,fall_formation_1980} is gradual, then so-called ``adiabatic contraction'' \citep{blumenthal_contraction_1986,gnedin_response_2004} will pull DM into the central regions, thereby increasing the central DM density and further steepening the slope of its density profile. It is thus natural to expect the maximum of the DM density to occur at the dynamical center of a galaxy, the lowest point of its gravitational potential. This expectation has made the Galactic Center (GC) a preferred target for indirect DM detection efforts searching for an annihilation signal \citep{bergstroem_observability_1998,gondolo_dark_1999,aharonian_hess_2006,baltz_pre-launch_2008,danninger_searches_2012}. 

As data from the Fermi Gamma-ray Space Telescope \citep{atwood_large_2009} has been accumulating, the number of studies reporting gamma-ray ``excesses'' or ``anomalies'' from the GC, that could be interpreted as a DM annihilation signal, has steadily grown \citep[see e.g.][]{goodenough_possible_2009,hooper_dark_2011,hooper_origin_2011,abazajian_detection_2012}. These analyses have searched for the broad gamma-ray continuum signal thought to arise from the decay and hadronization of DM annihilation products \citep{bergstroem_observability_1998}. While this is expected to be the dominant DM annihilation signature, it is unfortunately difficult to distinguish it from conventional sources (e.g. milli-second pulsars, Abazajian 2011\nocite{abazajian_consistency_2011}, or cosmic-ray interactions with molecular clouds, Yusef-Zadeh et al. 2012\nocite{yusef-zadeh_interacting_2012}), and at present a purely astrophysical explanation of these signals cannot be excluded.

More recently, however, there have been surprising reports of a highly statistically significant line-like feature at $\sim 130$ GeV in Fermi data from the GC \citep{bringmann_fermi_2012,weniger_tentative_2012,su_strong_2012}. Although such a DM annihilation line should be loop-suppressed by a factor $\sim 10^2 - 10^4$ compared to the expected continuum gamma-ray production in typical DM models, the fact that it is difficult to produce such a high energy line with astrophysical processes \citep[however, see][]{aharonian_cold_2012} has motivated further exploration of the DM annihilation explanation. In fact, DM particle physics models do exist in which the continuum is suppressed with respect to the line emission \citep[e.g.][]{cline_130_2012,buckley_implications_2012,bergstroem_130_2012,dudas_extra_2012,chalons_neutralino_2012}. In the analysis of \citet{su_strong_2012}, the line's significance is maximized at Galactic coordinates $(\ell,b) = (1.5^\circ,0^\circ)$, i.e. displaced from Sgr A*, the presumed dynamical center of the Galaxy, by about 200 projected parsec in the disk plane \citep[see also][]{tempel_fermi_2012}. This offset has been viewed as a strike against a DM annihilation interpretation of the line signal. It is possible that the offset is simply due to small number statistics \citep{yang_statistical_2012}, but nevertheless it is commonly viewed as a strike against a DM annihilation interpretation of the line signal.

On the other hand, the GC is a dynamically and energetically complex region \citep{genzel_galactic_2010}. It harbors a supermassive black hole (SMBH) \citep{ghez_measuring_2008,gillessen_monitoring_2009}, which may have been active as recently as 10 Myr ago, if the giant Fermi bubbles \citep{dobler_fermi_2010,su_giant_2010} are interpreted as resulting from a black hole accretion event; it is rich in massive young stars \citep{bartko_evidence_2009}, which likely formed in a single star burst event $6 \pm 2$ Myr ago \citep{paumard_two_2006}; it hosts plenty of highly energetic compact objects radiating in X-rays and gamma-rays \citep{muno_deep_2003,muno_overabundance_2005,abazajian_consistency_2011}; and on $\sim$kpc scales, its gravitational potential is non-axisymmetric due to the presence of a stellar bar and a boxy bulge \citep{blitz_direct_1991,martinez-valpuesta_unifying_2011}. Given that our Galaxy is baryon-dominated inwards of $\sim 5 - 10$ kpc \citep{klypin_cdm-based_2002}, one might expect astrophysical processes to modify the underlying DM distribution in significant ways. In principle these processes may even displace the maximum of the DM density away from the dynamical center, thus greatly affecting the expected DM annihilation signal from the GC.

In isolated galaxy simulations, resonant interactions between the stellar bar and the DM halo have been shown to alter the shape of DM halos reducing their triaxiality \citep{berentzen_stellar_2006,machado_loss_2010}, and to flatten a central cusp into a core \citep{weinberg_bar-driven_2002,athanassoula_what_2003,holley-bockelmann_bar-induced_2005,weinberg_bar-halo_2007}. The latter results remain controversial, however, since other numerical studies do not see such strong effects \citep{sellwood_bars_2003,valenzuela_secular_2003,colin_bars_2006}. Regarding an off-center DM density peak, these interactions are interesting because they may also induce a ``dark bar'' and other non-axisymmetric perturbations in the DM \citep{athanassoula_bar-halo_2002,ceverino_resonances_2007,mcmillan_halo_2005}. 

External gravitational perturbations, for example during a merger or a near passage of a satellite galaxy, could displace the tightly-bound baryonic component from the center of the overall mass distribution (the DM halo). Such offsets have been measured in galaxy clusters \citep{allen_resolving_1998}, in which separations between the center of the X-ray emitting gas and the gravitational center determined from strong lensing can be as large as $\sim 30$ kpc for relaxed clusters \citep{shan_offset_2010}. A study of THINGS galaxies by \citet{trachternach_dynamical_2008} found that offsets between the photometric and dynamical centers were less than one radio beam width ($\sim 10''$ or $150 - 700$ pc) for 13 out of 15 galaxies with well-constrained photometric centers. However, two galaxies in their sample, NGC 3627 (a barred Sb galaxy showing signs of a recent interaction) and NGC 6946 (a barred Scd galaxy), exhibit moderate offsets between one and two beam widths.

An offset DM density peak may be a reflection of an intrinsic lopsidedness in the Galaxy \citep{saha_milky_2009}. In fact, disk galaxies commonly exhibit substantial asymmetry in the central regions of their light distribution \citep[for a review, see][]{jog_lopsided_2009}. The origin of these asymmetries is not fully understood, with tidal encounters, gas accretion, and a global gravitational instability being some of the physical mechanisms under consideration. The central regions of advanced galaxy mergers often show long-lived unsettled sloshing behavior \citep{schweizer_colliding_1996,jog_measurement_2006}, and even in isolated galaxies the dynamical center can remain unrelaxed for many dynamical times \citep{miller_off-center_1992}, especially in systems with a cored mass distribution.

Supernova- or AGN-driven gas outflows may rapidly and non-adiabatically alter the potential in the central regions of proto-galaxies, prior to the formation of the bulk of their stars. Repeated episodes of such impulsive outflows, followed by slow adiabatic re-accretion of gas, may irreversibly transfer energy to the DM, flattening the central cusp in the process \citep{read_mass_2005,pontzen_how_2012}. Although this effect may not by itself produce an off-center DM density peak, the resulting cored density profile will be more susceptible to perturbations.

Lastly, the presence of a SMBH has been argued to lead to the formation of a steep cusp of DM \citep{gondolo_dark_1999,gnedin_dark_2004} in the inner parsec centered on the SMBH. This would obviously preclude any significant offset between the peak in DM annihilation signal and Sgr A*.

In this work we report on the search for the presence of an offset between the dynamical center and the maximum DM density in the Eris simulation \citep{guedes_forming_2011}, one of the highest resolution and most realistic cosmological simulations of the formation of a Milky-Way-like barred spiral galaxy. We find evidence for such an offset, at a scale of $300 - 400$ pc, consistent with the offset seen by \citet{su_strong_2012}. Since such an offset is only seen in the dissipational hydrodynamic simulations, and not in our collisionless pure-DM runs, we suggest that baryonic physics is in some way responsible for its formation. We examine a number of different physical mechanisms, but the limited resolution of this study does not allow us to conclusively settle on a single preferred explanation. At this stage, we wish to draw attention to the possibility of the maximum DM density (and hence annihilation luminosity) not being coincident with the dynamical center of our Galaxy, commonly associated with Sgr A*. We hope that our results will stimulate future work, examining other high resolution hydrodynamic galaxy formation simulations, and investigating in more detail the physical mechanisms that can give rise to an offset DM density peak.

The remainder of this paper is organized as follows. In Section 2 we describe the numerical simulations that we have analyzed. In Section 3 we present the evidence for a DM offset in the Eris simulation. In Section 4 we go over several possible formation mechanisms and confront each of them with data from the simulations. In Section 5 we discuss implications for indirect detection searches towards the GC, and finally in Section 6 we present our conclusions.

\section{Simulations}

\begin{deluxetable*}{lccccccc}
\tablecaption{Four Cosmological Zoom-in Simulations.}
\tablehead{
\colhead{Name} & \colhead{Code} & \colhead{$m_{\rm DM} \; [{\rm M_\odot}]$} & \colhead{$\softening \; [{\rm pc}]$}   & \colhead{$N_{\rm vir}$} & \colhead{$M_{\rm vir} \; [{\rm M_\odot}]$} & \colhead{$R_{\rm vir} \; [{\rm kpc}]$} & \colhead{$(\alpha, \, r_{-2} \; [{\rm kpc}], \, \rho_{-2} \; [\msun \, {\rm kpc}^{-3}])$}
}
\startdata
Eris & Gasoline & $9.8 \times 10^4$ & 124 & $1.9 \times 10^7$ & $7.9 \times 10^{11}$ & 239 & $(0.0984, \; 18.8, \; 3.84 \times 10^6)$ \\
ErisDark & Gasoline & $1.2 \times 10^5$ & 124 & $7.6 \times 10^6$ & $9.1 \times 10^{11}$ & 247 & $\;\;(0.173, \; 42.5, \; 8.73 \times 10^5)$ \\
Via Lactea II & PKDGRAV2 & $4.1 \times 10^3$ & 40 & $4.1 \times 10^8$ & $1.7 \times 10^{12}$ & 309 & $\;\;(0.144, \; 53.0, \; 7.78 \times 10^5)$ \\
GHalo & PKDGRAV2 & $1.0 \times 10^3$ & 60 & $1.1 \times 10^9$ & $1.1 \times 10^{12}$ & 267 & $\;\;(0.137, \; 61.1, \; 3.53 \times 10^5)$
\enddata
\tablecomments{$m_{\rm DM}$ is the mass of high-resolution DM particles, \softening\ their gravitational softening. $N_{\rm vir}$ and $M_{\rm vir}$ are the total number and mass of all particles (including gas and star particles in Eris) within the virial radius $R_{\rm vir}$, defined as the radius enclosing a density of 92.5 times the critical density. The last column lists the parameters of the best-fit Einasto profile ($\ln( \rho / \rho_{-2}) = -2/\alpha \, [(r/r_{-2})^\alpha - 1]$) for the mean enclosed density $\langle \rho \rangle(<\!r)  = $ $M(<\!r)/(4\pi/3 \, r^3)$ (see Fig.~\ref{fig:density_profiles}). \label{tab:simulations}}
\end{deluxetable*}

Our analysis makes use of four cosmological zoom-in simulations of the formation and evolution of a Milky-Way-analog galaxy: one DM+hydrodynamics simulation, ``Eris'' \citep{guedes_forming_2011}; and three pure-DM dissipationless simulations, ``ErisDark'' (Pillepich et al., in prep.), ``Via Lactea II'' \citep{diemand_clumps_2008}, and ``GHalo'' \citep{stadel_quantifying_2009}. Details of these simulations are presented in the listed references, and we only briefly summarize their salient features here. Eris and its DM-only twin ErisDark were run with the N-body+SPH code \textit{Gasoline} \citep{wadsley_gasoline:_2004}, but only Eris utilized the SPH hydrodynamics. Eris resolves the formation and evolution of a galaxy with $1.3 \times 10^7$ high resolution DM particles of mass $9.8 \times 10^4 \msun$ and a similar number of SPH gas particles with an initial mass of $1.2.1 \times 10^4 \msun$. Gravitational interactions are softened using a cubic spline density kernel of length $\softening = 124$ proper pc\footnote{Forces become Newtonian at 2 \softening.}. Throughout the course of the simulation, star particles are created from dense gas ($> 5$ atoms cm$^{-3}$) according to a heuristic star formation recipe with a 10\% star-formation efficiency. These star particles affect surrounding gas through a supernova blastwave feedback prescription that injects thermal energy, mass, and metals. The simulation results in a realistic looking barred late-type spiral disk galaxy, that matches many observational constraints on the structure of the Milky Way. For example, it has a low bulge-to-disk ratio of 0.35, falls on the Tully-Fisher relation, has a stellar-to-total mass ratio of 0.04, and a star formation rate of 1.1 M$_\odot$ yr$^{-1}$. Eris is the most realistic such simulation available today. 

ErisDark is a DM-only twin to Eris, meaning that it was initialized with the same phases of the Gaussian random density field, but all of the matter is treated as DM, while in Eris 17\% is baryonic. ErisDark thus has a slightly higher DM particle mass of $1.2 \times 10^5 \msun$, but employs the same \softening. A detailed comparison of the two simulations will be presented in Pillepich et al. (2012, in preparation). Lastly, we also compare results with two of the highest resolution pure-DM simulations ever performed, Via Lactea II \citep[VL2,][]{diemand_clumps_2008} with a particle mass of $m_p = 4,100 \msun$ and a force softening of $\softening = 40$ pc and GHalo \citep{stadel_quantifying_2009} with $m_p = 1,000 \msun$ and $\softening = 60$ pc. Both VL2 and GHalo were run with the purely collisionless N-body code PKDGRAV2. The relevant parameters of all simulations are summarized in Table~\ref{tab:simulations}.

\section{An Off-Center DM Density Peak}

We define the dynamical center of the halo to be the location of the minimum of the total gravitational potential. The gravitational potential is calculated during the simulation by solving the Poisson equation, with source terms contributed by all DM, star, and gas particles \citep[for details, see][]{wadsley_gasoline:_2004}. In the Eris simulation the potential minimum is typically set by the stars, which dominate the potential in the center of the galaxy. The location of the potential minimum is coincident with the center of mass of the stellar disk (and with the point of maximum stellar and gas density) to within 10 pc. We select for our analysis all particles within a radius of 2.5 kpc from the dynamical center. At $z=0$ there are 104,781 DM, 13,792 gas, and 4,597,762 star particles in this region in Eris, contributing $1.03 \times 10^{10}$, $2.07 \times 10^8$, and $1.97 \times 10^{10} \msun$ in mass. In ErisDark, VL2, and GHalo there are 42,611, 1,718,223, and 4,041,566 DM particles in this region, for a mass of $5.06 \times 10^9$, $7.04 \times 10^9$, and $4.12 \times 10^9 \msun$, respectively.

\subsection{Enclosed density profiles}

\begin{figure}
\centering
\includegraphics[width=\columnwidth]{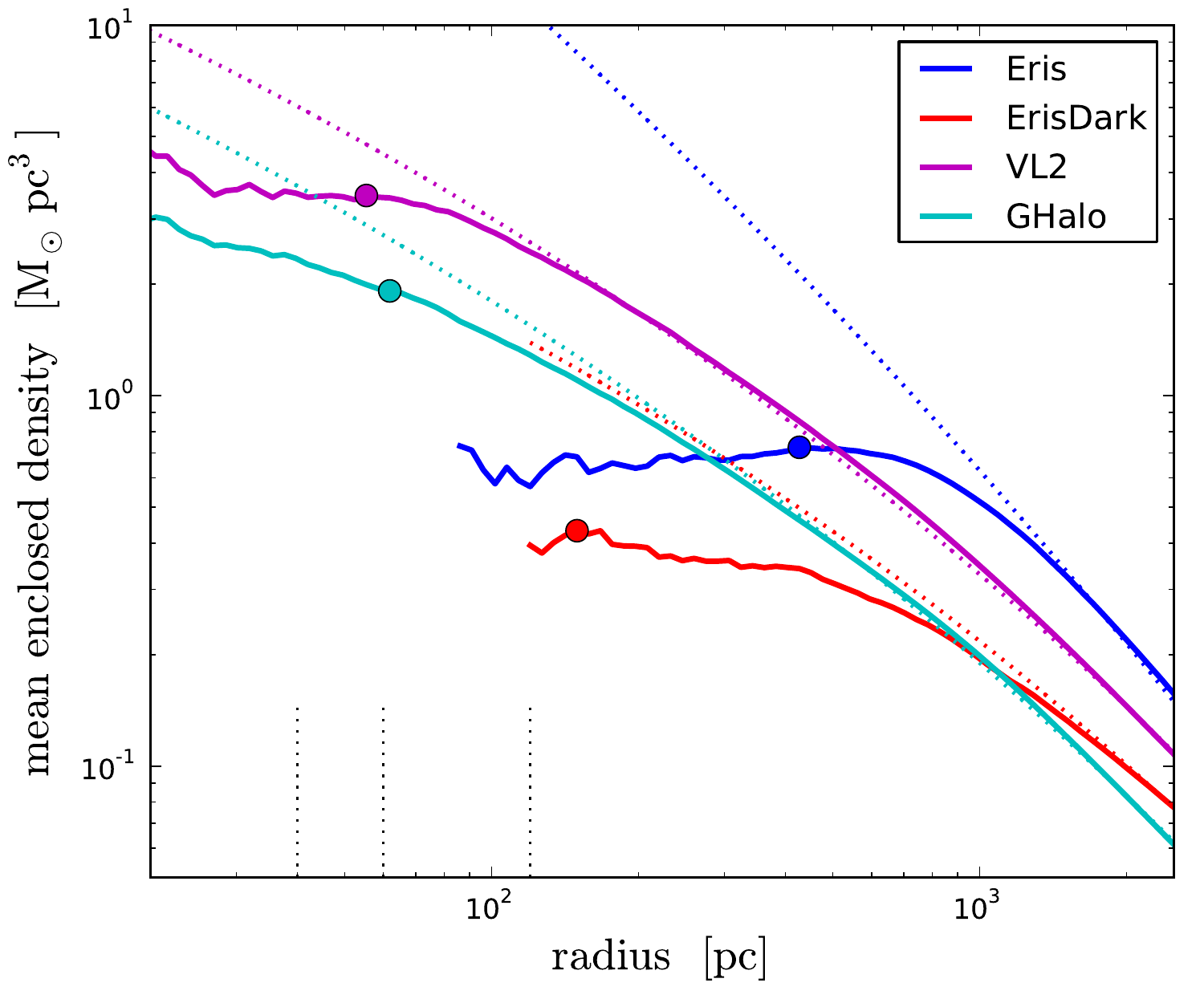}
\caption{Radial profiles of the average enclosed dark matter density $\langle \rho \rangle(<\!r)$ centered on the location of the potential minimum, starting at the radius enclosing at least 20 particles. The vertical dotted lines indicate the softening lengths, \softening\ = 124, 124, 40, and 60 pc for Eris, ErisDark, VL2, and GHalo, respectively. The location of the maximum enclosed density (for $r > \softening$) is denoted with a filled circle. Dotted lines show the best-fit Einasto profiles (fitted to 1 kpc $< r < R_{\rm vir}$), whose parameters are given in Table~\ref{tab:simulations}.}
\label{fig:density_profiles}
\end{figure}

Fig.~\ref{fig:density_profiles} shows radial profiles of the averaged enclosed DM density, $\langle \rho(<\!r) \rangle = M(<\!r)/(4\pi / 3 r^3)$, for the four simulations. The densities are higher in VL2 than in GHalo and ErisDark owing to the larger virial mass and concentration of the VL2 host halo, but in Eris even higher DM densities are reached at $r > 500$ pc due to the baryonic contraction. Compared to ErisDark the density profile is slightly steeper in Eris, and this is also reflected in a much higher halo concentration. At even smaller radii, baryonic physics produces a roughly constant mean density core.\footnote{These density profiles are centered on the potential minimum. Centering on the point of maximum DM density slightly changes the central slope.} The radii at which the enclosed density is maximized are denoted with filled circles in Fig.~\ref{fig:density_profiles}, and it is clear that in Eris this maximum is significantly offset from the dynamical center, while in the dissipationless simulations it occurs at close to one \softening.

\subsection{Three-dimensional localization of the density peak}

\begin{figure}
\centering
\includegraphics[width=\columnwidth]{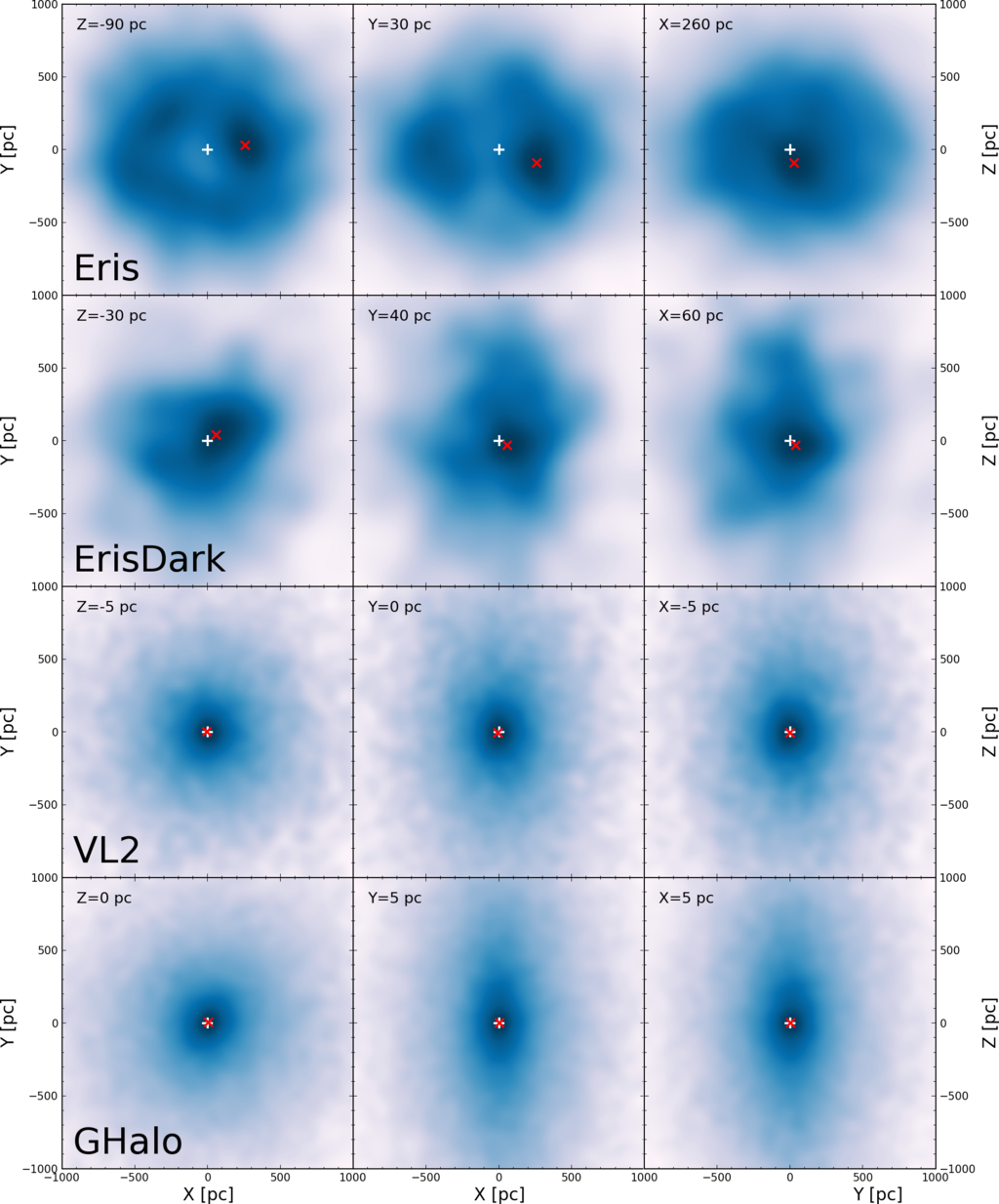}
\caption{Slices of the central 2 kpc $\times$ 2 kpc DM density field at $z=0$ through the location of maximum DM density. From left to right: x-y, x-z, and y-z planes; from top to bottom: Eris, ErisDark, VL2, and GHalo simulations. The position of the slice in the perpendicular direction is given in the top left corner of every panel. The density field has been smoothed with a Gaussian kernel of width $\sigma$ equal to one gravitational softening length \softening\ = 124, 124, 40, and 60 pc, respectively. The z-axis coincides with the disk normal in Eris, and with the major axis of the prolate DM density ellipsoid in the other cases. The slice thickness is 10pc for Eris and ErisDark and 5pc for VL2 and GHalo. The images are centered on the minimum of the total potential (marked with a white cross), while the location of the maximum density is indicated with a red 'x'. In the three dissipationless simulations (ErisDark, VL2, and GHalo) the offset between the maximum DM density and the total potential minimum is less than \softening, but it is 2.3 \softening\ in Eris.}
\label{fig:slice_all}
\end{figure}

To study the DM density offset in more detail, we deposit DM particles using a cloud-in-cell algorithm onto a three-dimensional grid with cell width equal to 10 pc for Eris and ErisDark ($500^3$ grid) and 5 pc for VL2 and GHalo ($1000^3$ grid). The grid is oriented in space such that the z-axis is aligned with the disk normal in Eris, and with the major axis of the prolate density ellipsoid for the other cases. We smooth the discretized 3D density fields by applying a Gaussian smoothing kernel of width $\sigma$. Fig.~\ref{fig:slice_all} shows slices through this grid with $\sigma=\softening$. The position of the cell with the maximum DM density is indicated with a red cross, and again it is clear that it is significantly displaced from the dynamical center is Eris. The maximum occurs at $(x,y,z) = (260, 30, -90)$ pc, corresponding to an offset distance $\Doff = 280$ pc, which is about 2.3 \softening. In the dissipationless simulations on the other hand, $\Doff = 78$, 7.1, and 7.1 pc, all well within one \softening.

\begin{figure}
\centering
\includegraphics[width=\columnwidth]{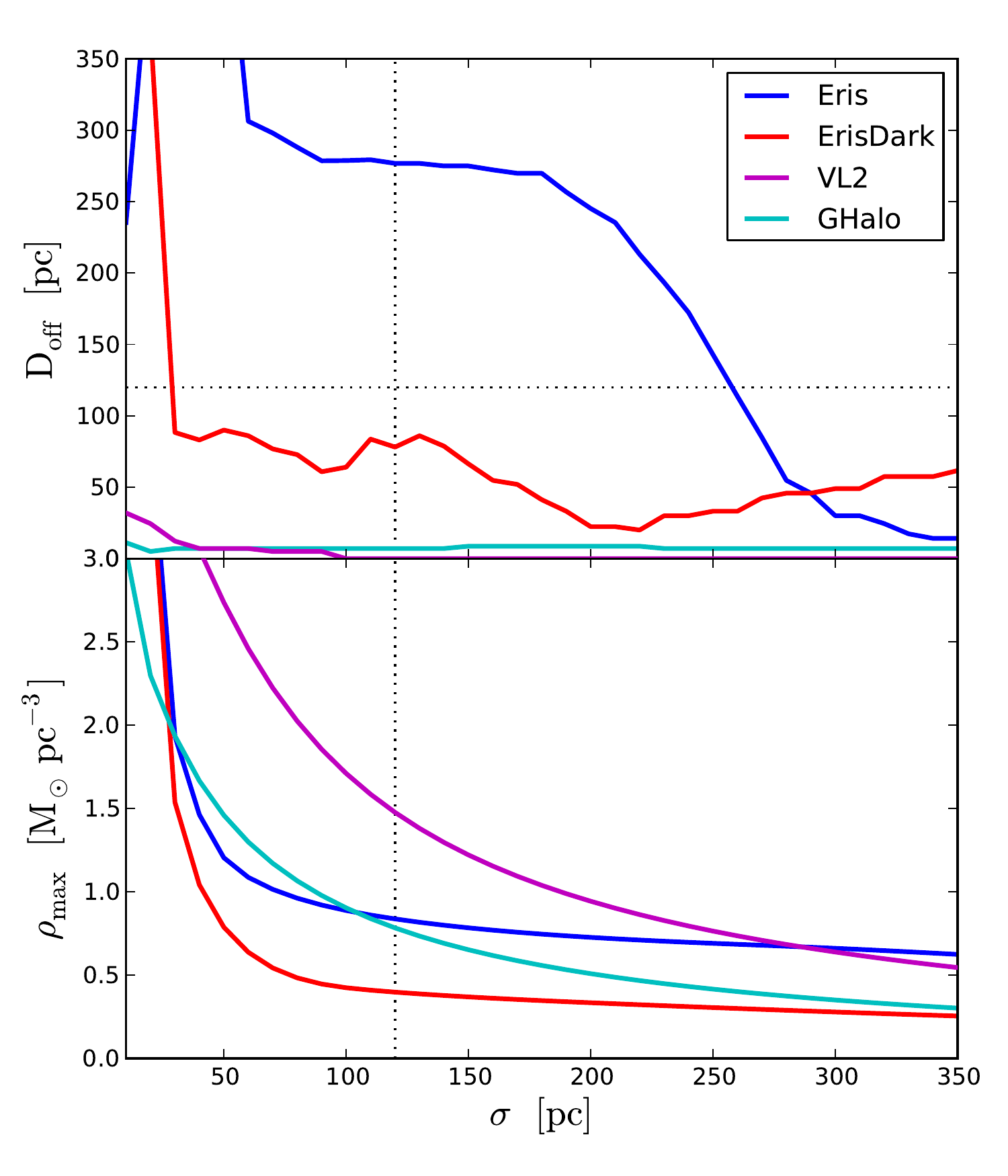}
\caption{The dependence of \Doff\ and $\rho_{\rm max}$ on the width $\sigma$ of the Gaussian density smoothing kernel. The gravitational softening scale $\epsilon_{\rm soft} = 124$ pc for Eris and ErisDark is indicated with the dotted line. In the dissipationless simulations (ErisDark, VL2, and GHalo) the DM offset remains well below \softening\ for all $\sigma$. In Eris the offset is larger than \softening\ out to $\sigma=260$ pc (about 2 \softening). The maximum density is about a factor of two higher in Eris than in ErisDark, and depends only mildly on $\sigma$ for $\sigma > \softening$. $\rho_{\rm max}$ is larger in VL2 due to its higher halo mass.}
\label{fig:offset_vs_sigma}
\end{figure}

We have repeated the determination of $\Doff$ for different choices of $\sigma$, and the results are shown in the top panel of Fig.~\ref{fig:offset_vs_sigma}. The values of \Doff\ for $\sigma \ll \softening$ are not meaningful, since shot noise can lead to unphysical density spikes on very small scales. But for $\sigma \gtrsim \softening$ the conclusions are robust: the DM offset remains below one \softening, i.e. is consistent with no offset, for all three DM-only simulations regardless of smoothing, while for Eris the offset is greater than \softening\ up to $\sigma \approx 260$ pc. That $\Doff$ drops to zero for even higher $\sigma$ is not surprising, since any density distribution will appear centered when smoothed on sufficiently large scales.

In the bottom panel of Fig.~\ref{fig:offset_vs_sigma} we show the maximum value of the density, $\rho_{\rm max}$, for our four simulations. Of course $\rho_{\rm max}$ is much more sensitive to $\sigma$ than \Doff, since larger smoothing includes contributions from surrounding lower density material. At $\sigma=\softening$ the density peaks a value of $0.84 \msun \, {\rm pc}^{-3}$ in Eris, about 25\% higher than the value ($0.67 \msun \, {\rm pc}^{-3}$) at its dynamical center. In ErisDark $\rho_{\rm max}=0.40 \msun \, {\rm pc}^{-3}$, less than half of the peak density in Eris. We caution, that the resolution of our numerical simulations is not sufficient to resolve the internal density structure of the offset peak. With higher resolution run the peak-to-center density contrast may well be higher.

\subsection{Evolution of the DM offset}\label{sec:evolution}

\begin{figure}
\centering
\includegraphics[width=\columnwidth]{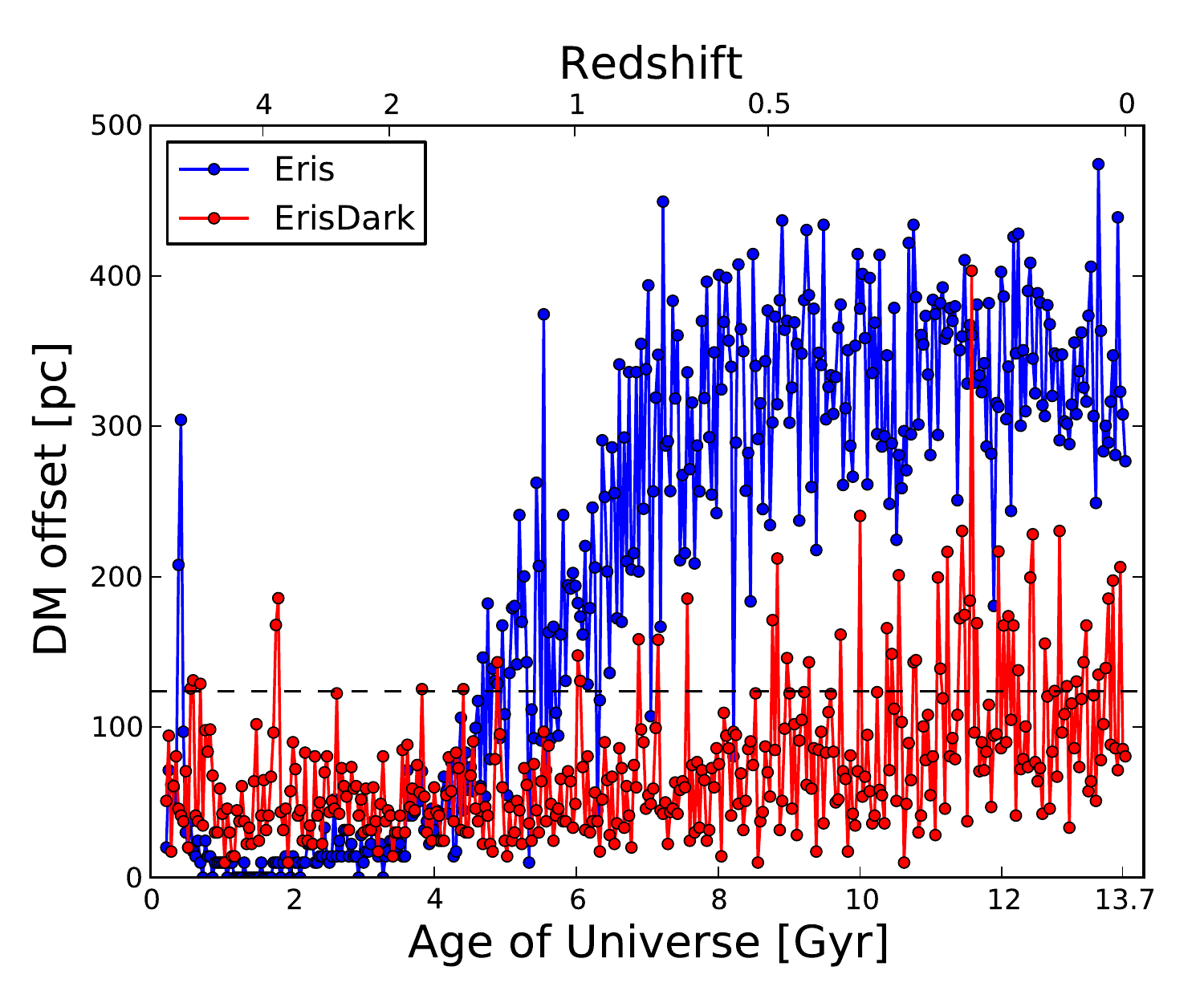}
\caption{The time dependence of the DM offset from the total potential minimum in Eris (\textit{blue}) and ErisDark (\textit{red}). The gravitational softening $\softening(z)=124$ pc is indicated with a dashed line. In ErisDark the DM offset remains around or below 1 \softening\ for almost the entire simulation, while in Eris it begins to significantly exceed \softening\ around $z=1.5$ and remains at $\approx 3 \softening$ afterwards.}
\label{fig:offset_evolution}
\end{figure}

We have performed the same analysis described in the previous section on all 400 outputs of the Eris and ErisDark simulations. Fig.~\ref{fig:offset_evolution} shows that the offset measured in Eris at $z=0$ is no fluke, but persists over cosmological time scales. At very early times ($z \gtrsim 2$) there is no offset in either Eris or ErisDark. Starting at $z \approx 1.5$, however, the DM density maximum in Eris starts to depart from the dynamical center. Over a period of about two Gyr the DM offset grows to $\Doff \approx 340$ pc (almost 3 \softening), where it remains for the remainder of the simulation. In contrast, in the ErisDark simulation \Doff\ remains below 1 \softening\ for almost its entire evolution, albeit with occasional spikes up to $\sim 200$ pc. These results are qualitatively quite similar to those reported by \citet[see their Fig.4]{maccio_halo_2012} in a similar, albeit 8 times lower resolution, simulation.

In Eris, \Doff\ fluctuates around its time average (over the last 4 Gyr) of 340 pc with a root mean square (rms) dispersion of 51 pc. The closest the peak comes to the center over this time is $\Doff = 180$ pc. The peak preferentially lies near the disk plane; its mean vertical (perpendicular to the disk plane) displacement is only $\langle |z| \rangle = 64$ pc, with an rms dispersion of 46 pc. The maximum density varies around a mean value of $\langle \rho_{\rm max} \rangle = 0.84 \msun\ {\rm pc}^{-3}$ with an rms dispersion of $0.02 \msun\ {\rm pc}^{-3}$, and reaches minimum and maximum values of 0.79 and 0.92 $\msun\ {\rm pc}^{-3}$.

Outputs in the Eris simulation are spaced $\sim 35$ Myr apart, which is too long to resolve the dynamics of the offset peak, given the local dynamical time of $\sim 15$ Myr. Nevertheless it is already clear from looking at this coarsely time-sampled data that the peak locations are not randomly distributed throughout the central region. We defer further discussion of the temporal evolution of the DM offset and implications for its physical nature to Section~\ref{sec:origin}.

\begin{figure}
\centering
\includegraphics[width=\columnwidth]{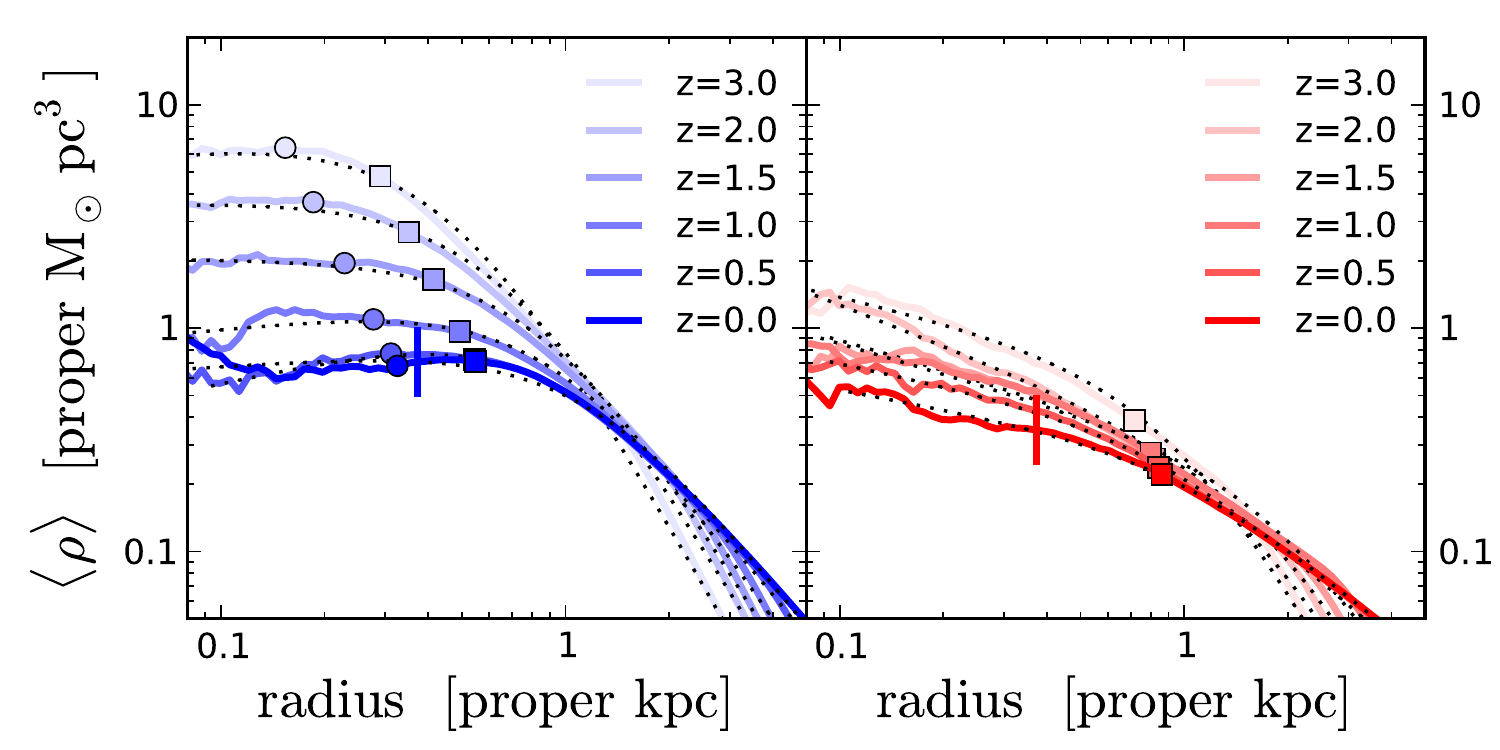}
\includegraphics[width=\columnwidth]{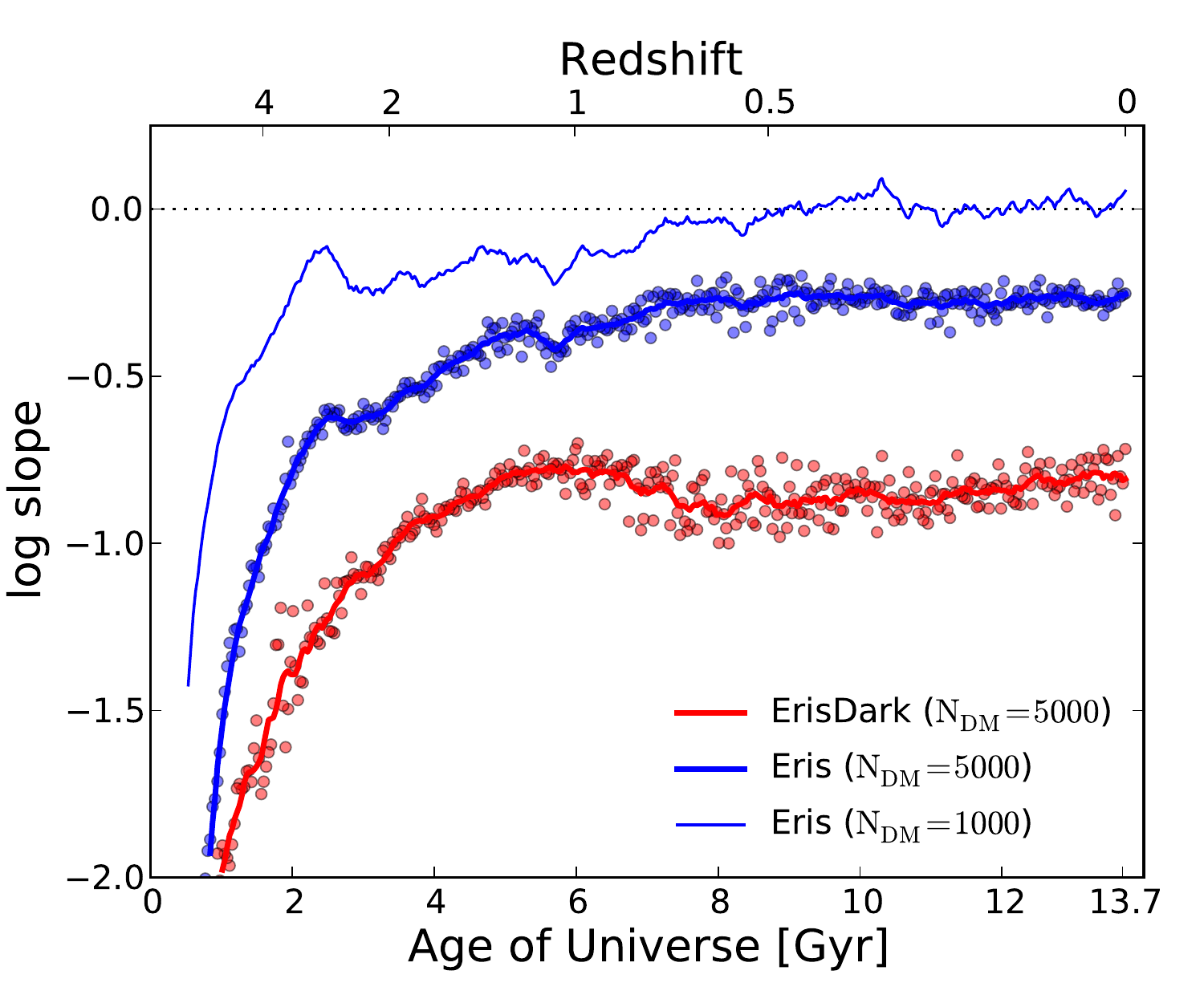}
\caption{Evolution of the inner DM density profiles. \textit{Top:} Density profiles at $z=$ 3, 2, 1.5, 1, 0.5, and 0 for Eris (left) and ErisDark (right). All quantities are plotted in proper units. The black dotted lines indicate the best-fitting modified Burkert profile (see Eq.~\ref{eq:modified_Burkert}). The location of $r_{5000}$, the radius enclosing 5000 DM particles, is indicated with squares. $r_{5000}$ corresponds to the density profile convergence radius in ErisDark at $z=0$. For Eris the density profile can be trusted to smaller radii due to its larger particle counts, and so we additionally mark $r_{1000}$ with a circle. The vertical bars indicate $r=3\softening$. \textit{Bottom:} The evolution of the logarithmic slope ($d\!\ln\rho/d\!\ln r$) measured at different radii: the symbols and thick lines correspond to the slopes at $r_{5000}$, and for Eris we also show the slope at $r_{1000}$. All lines are boxcar averages over 10 outputs. This plot shows that baryonic physics in Eris leads to the flattening of the central density profile. This flattening appears to be correlated with the growth of the DM offset (cf. Fig.~\ref{fig:offset_evolution}). We emphasize that \textit{these slopes are not asymptotic slopes}, and a value of less than $-1$ does not imply a strongly cusped profile all the way to the center, but merely indicates the local slope at $r_{5000}$, which may lie outside of the scale radius at early times.}
\label{fig:profile_evolution}
\end{figure}

\subsection{Correlation with DM density profile flattening} \label{sec:DM_core}

The growth of the DM density offset appears to be well correlated with a flattening of the central DM density profile in Eris, as demonstrated in Fig.~\ref{fig:profile_evolution}. The top panels show mean enclosed density profiles in the inner region ($r < 5$ proper kpc) at several output times from $z=3$ to $z=0$. We see two notable differences between Eris (left panel) and ErisDark (right). First, the DM densities (plotted in proper units) tend to be higher in Eris than in ErisDark, which is indicative of ``adiabatic contraction''. Secondly, while the enclosed density in ErisDark continues to increase towards smaller radii, the profiles flatten out in Eris, indicating the formation of a core. Note that we use the term ``core'' loosely, indicating a substantial flattening of the density profile, but not necessarily implying a constant density.

Inner density profiles are notoriously difficult to properly resolve in N-body simulations, requiring high force resolution, large particle counts, and accurate time integration with sufficiently small timesteps \citep{power_inner_2003,zemp_optimum_2007,diemand_clumps_2008,dubinski_anatomy_2009}. In Pillepich et al. (in preparation) we investigate the influence of baryons on the full shape of the density profiles in more detail, including convergence studies. We find a convergence radius for the $z=0$ density profile in ErisDark of 0.9 kpc, corresponding to $\sim 5000$ enclosed particles. The square symbols in the top panel of Fig.~\ref{fig:profile_evolution} indicate $r_{5000}$ (the radius enclosing 5000 DM particles) for the different outputs. Note that $r_{5000}$ is significantly farther out than $3 \, \softening$ (indicated with the vertical bars), a commonly used ``rule of thumb'' for how far in density profiles can be trusted. For Eris, $r_{5000}$ is probably overly conservative, since the potential is dominated by baryons and many more than 5000 gas and star particles are found inside of $r_{5000}$. As an intermediate case we also mark with circles $r_{1000}$, the radius enclosing 1000 DM particles.\footnote{At $z=0$, there are over a million star particles within $r_{5000}$ and more than 400,000 within $r_{1000}$.}

In the bottom panel of Fig.~\ref{fig:profile_evolution} we show the evolution of the logarithmic slope $d\!\ln \rho / d\!\ln r$, measured at these radii. We obtained these slopes by first fitting the enclosed density profiles inwards of 5 kpc to a modified Burkert profile,
\begin{equation}
\langle \rho \rangle(r) = \frac{\rho_0 \, r_c^{\beta}}{r^\alpha \, \left(r^2 + r_c^2\right)^{(\beta-\alpha)/2}}, \label{eq:modified_Burkert}
\end{equation}
and then evaluating its logarithmic slope at the radii of interest,
\begin{equation}
\frac{{\rm d}\!\ln\langle \rho \rangle}{{\rm d}\!\ln r} = -\alpha + \frac{\alpha - \beta}{1 + (r_c/r)^2}.
\end{equation}
This functional form describes the central enclosed density profile in Eris and ErisDark much better than a constant power law or Einasto profile. Note, however, that this fitting function is designed to track the enclosed density profile all the way down to \softening, so \textit{way beyond the convergence radius}. We wish to emphasize that we merely use this fit to evaluate the logarithmic slope at radii outside of 3 \softening, and do not attach any physical meaning to the asymptotic inner slopes $\alpha$ preferred by the fits.

Fig.~\ref{fig:profile_evolution} clearly shows that Eris develops a cored density profile, in the sense that it has a much shallower logarithmic slope at $r_{5000}$ than ErisDark. Measured at $r_{5000}$ the log slope in Eris becomes progressively shallower until it stabilizes at a value of $\sim -0.3$ at $z \approx 1$. Note that $r_{5000}$ is larger than $\Doff(z)$ at all redshifts. Measured at $r_{1000}$ the profile even flattens out completely with a log slope of $\sim 0$. In ErisDark, on the other hand, the slope at $r_{5000}$ remains close to the NFW value of $-1$ after $z=1.5$. Comparison with Fig.~\ref{fig:offset_evolution} shows that the onset and time scale of the core formation in Eris is remarkably well correlated with the growth of the DM density offset. This is a strong hint that whatever baryonic process drives the core formation may also be responsible for the offset DM density peak.

\section{Nature and Origin of the Offset Peak}\label{sec:origin}

The existence of a well defined offset to the peak of the DM distribution in Eris is unexpected, and its physical nature not immediately apparent. Some possibilities that we have considered are: (i) statistical fluctuations, (ii) an incompletely dissolved subhalo core, and (iii) excitation of a DM density wave by the stellar bar or other perturber. In the following sections we address these possibilities in turn.

\begin{figure}
\centering
\includegraphics[width=\columnwidth]{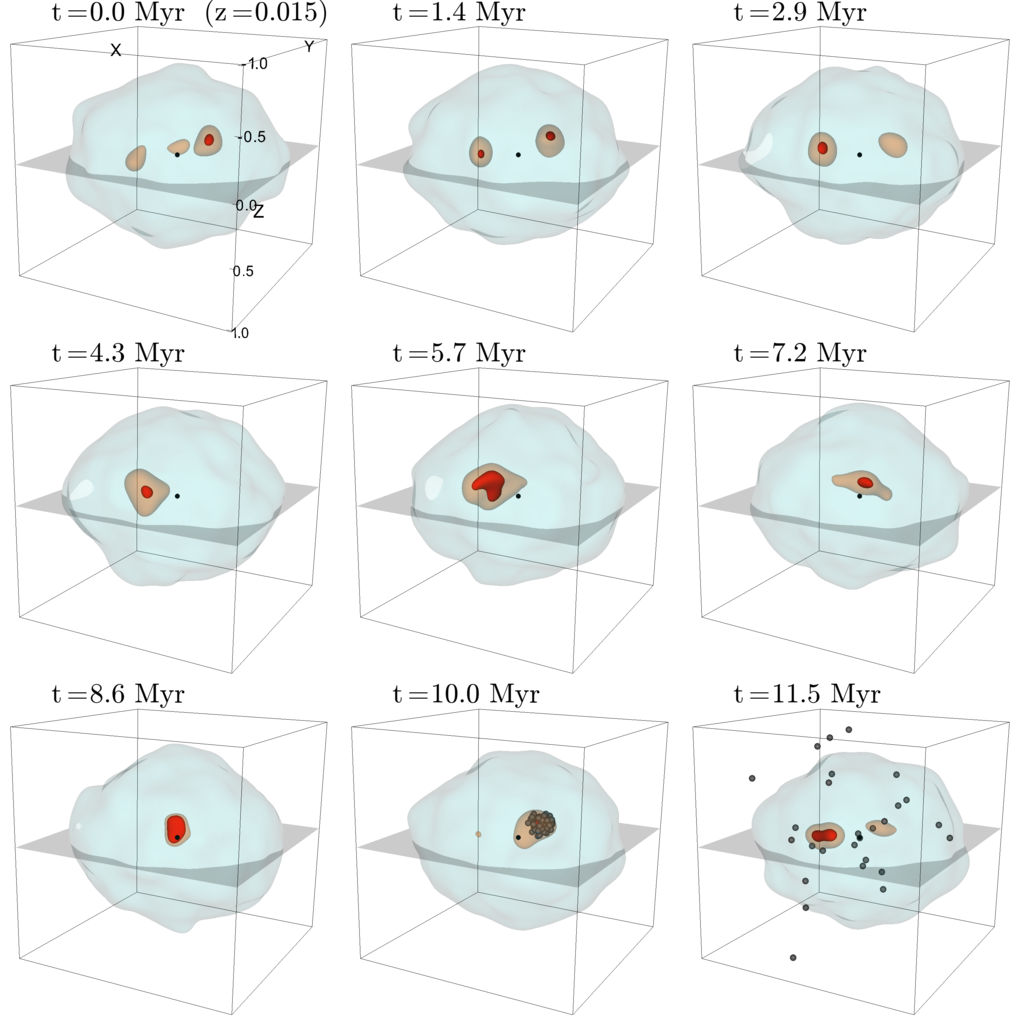}
\caption{3D contour plots of the DM density in the central $(2 \, {\rm kpc})^3$ volume of Eris. The outer contour (light blue) corresponds to $\rho = 0.45 \, \msun\ {\rm pc}^{-3}$, the middle contour (light orange) to $0.8  \, \msun\ {\rm pc}^{-3}$, and the opaque innermost contour (red) to the 99$^{\rm th}$ percentile of the DM density in the volume. The dynamical center is marked with a black dot, and the stellar disk lies in the X-Y plane. The images show the time evolution over 11.5 Myr ($\Delta t = 1.43$ Myr) starting from $z=0.0151$. In the last two frames we have plotted the positions of all 78 particles located within 1 \softening\ of the maximum density at the second-to-last output shown. Only 1.43 Myr later (last frame) they have already dispersed throughout the plotted volume, indicating that the density offset is not a bound structure.}
\label{fig:high_cadence_9panel}
\end{figure}

As we have seen, the output cadence of the original Eris simulation is not sufficient to follow the dynamics of the peak. To help with understanding the physical nature of the offset, much higher cadence outputs are desirable. To this end we have restarted the Eris simulation from output number 393 ($z=0.0151$) with a $\sim 20$ times finer temporal resolution ($\Delta t = 1.43$ Myr) and evolved the simulation for 140 time steps (206 Myr). Visualizations of nine consecutive high cadence outputs, spanning 11.5 Myr, are shown in Fig.~\ref{fig:high_cadence_9panel}, and a movie of the full evolution can be viewed at \texttt{\small http://vimeo.com/45114776}.

Visual inspection of these much higher cadence outputs shows that the offset peak appears to sometimes jump around somewhat discontinuously between two subsequent outputs. Occasionally there are multiple peaks of roughly equal density. As already seen with the coarser time resolution, the point of maximum density appears to avoid the very center and remains close to the disk plane.

\begin{figure}
\centering
\includegraphics[width=\columnwidth]{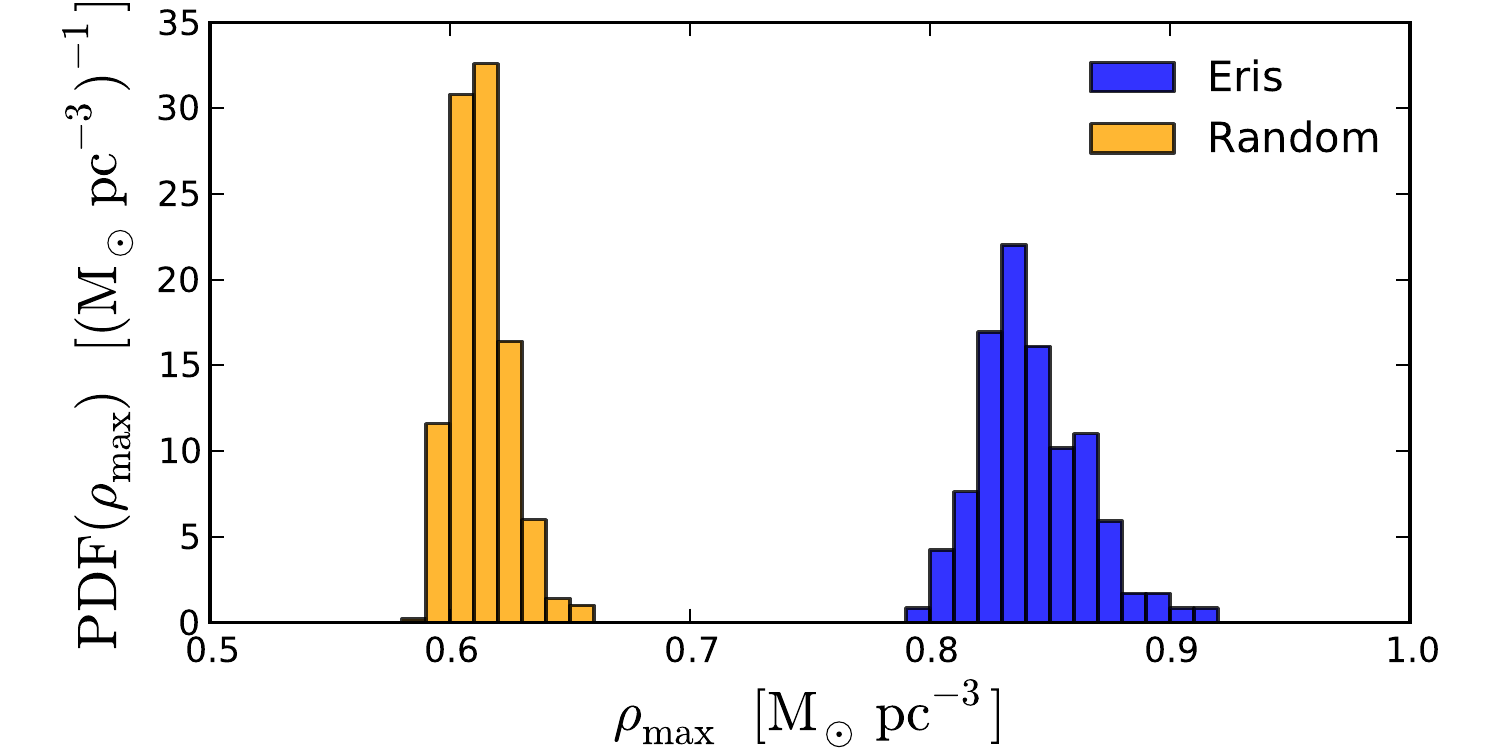}
\includegraphics[width=\columnwidth]{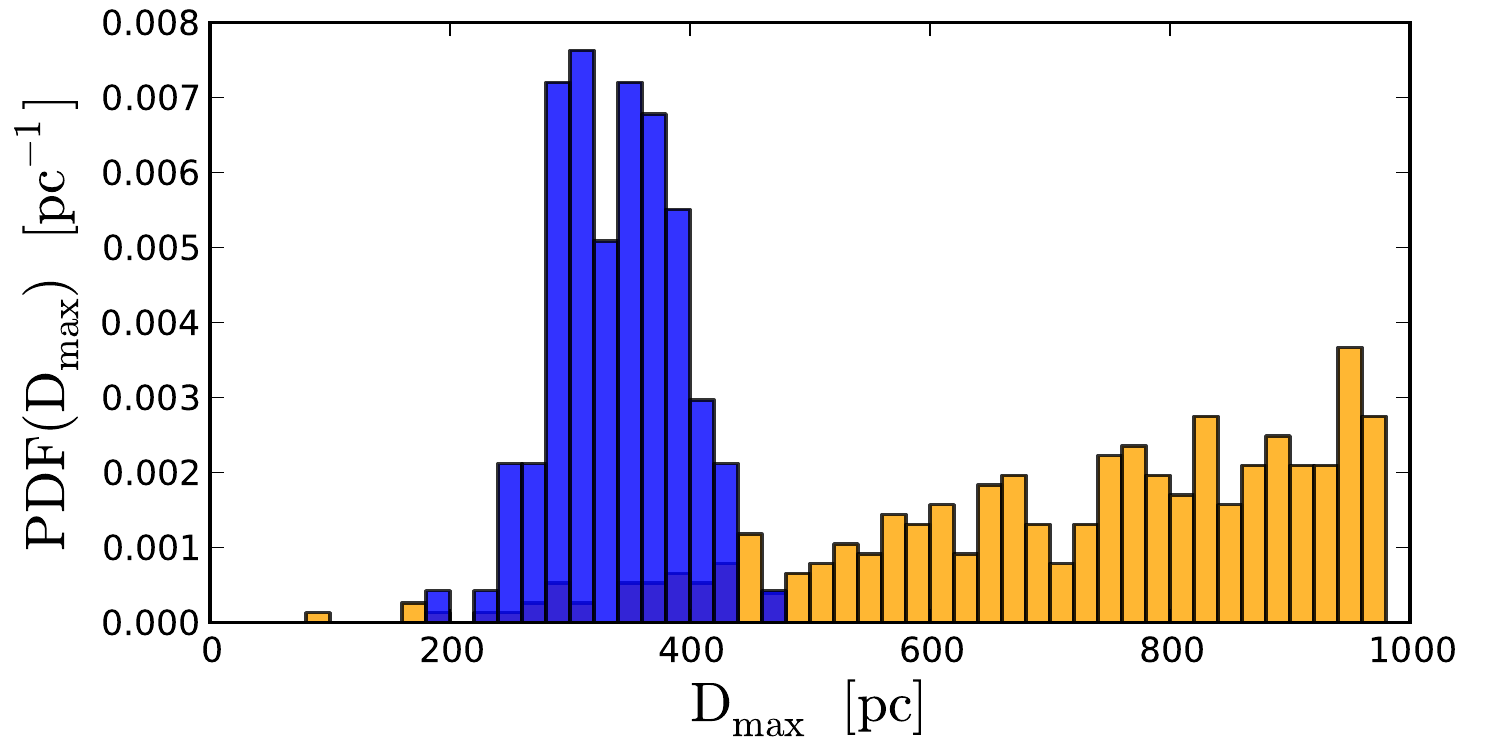}
\caption{Probability distribution functions of the maximum density $\rho_{\rm max}$ (\textit{top panel}) and its distance to the potential minimum \Doff\ (\textit{bottom}), compared between the Eris simulation (\textit{blue}) and random samples (\textit{orange}). The Eris sample consists of the last $\sim 124$ outputs spanning 4 Gyr of evolution. The random sample is comprised of 500 realizations of a randomly drawn particle distributions with the same uniform mean density as the Eris core inwards of 1 kpc.}
\label{fig:fluctuations}
\end{figure}

\subsection{Statistical fluctuations?}
As shown in Sec.~\ref{sec:evolution}, the growth of the DM offset coincides in time with the formation of a core in the Eris DM distribution. In a constant density core, sampled with a small number of N-body particles, Poisson fluctuations can give rise to a spurious density peak offset from the center. The mean DM density in the Eris' core (inwards of 1 kpc) is $0.51 \msun \, {\rm pc}^{-3}$, corresponding to an N-body particle density of 5,300 particles kpc$^{-3}$. At $z=0$ there are 509 particles within 2 \softening\ of the location of maximum density, much higher than the expected 306 for the average core density and well beyond what Poissonian fluctuations can produce. We have numerically confirmed this by constructing 500 randomly drawn particle distributions with the same mean particle density as the Eris core, and analyzed these samples in the same manner as described above.

In Fig.~\ref{fig:fluctuations} we compare the distributions of $\rho_{\rm max}$ and \Doff\ between the random samples and the last $\sim 124$ Eris outputs spanning 4 Gyr of evolution. Eris' $\rho_{\rm max}$ distribution is peaked at $0.85 \msun \, {\rm pc}^{-3}$, way beyond the random distribution, which peaks at $0.61 \msun \, {\rm pc}^{-3}$ and has an rms dispersion of only $0.01 \msun \, {\rm pc}^{-3}$. Similarly, the \Doff\ probability distribution function for Eris is peaked around 350 pc and looks nothing like that of the random samples, which slowly increases towards larger \Doff, in accordance with a simple volumetric $\Doff^{1/3}$ scaling. These comparisons very clearly demonstrate that the DM offsets we have found in Eris are not due to statistical fluctuations.

\subsection{Incompletely disrupted subhalo core?}

In this potential explanation, the offset density peak would be identified with the tightly bound central regions of an incompletely disrupted subhalo. The subhalo would have to have been massive enough to have experienced substantial baryonic condensation and associated contraction of its DM, making it more resilient to complete tidal disruption than its DM-only counterpart in ErisDark. If such a halo fell into the host at $z \gtrsim 1.5$, dynamical friction would quickly drag it in to the center, where tidal interactions would have stripped most of the weakly bound material in its outskirt. The inspiral would have stalled once its mass dropped to the point where dynamical friction becomes unimportant. After that point it would continue to orbit around the dynamical center, its temporal cohesion possibly aided by the stabilizing effect of a harmonic potential \citep{kleyna_dynamical_2003,read_dynamical_2006}.

We can rule out this possibility, based on several observations: (i) as mentioned above there is no stellar counterpart to the DM offset, and (ii) the peak persists for hundreds of dynamical times, which seems too long for an orbiting subhalo core to survive. The final nail in the coffin (iii) is the fact that the offset DM peak does not appear to be a bound feature. This is demonstrated in the last two panels of Fig.~\ref{fig:high_cadence_9panel}, in which we show the positions of all 78 particles located within 1 \softening\ of the maximum density peak at $t=10$ Myr (after $z=0.015$). A mere 1.4 Myr later, these same particles have dispersed throughout the central volume. This implies that the offset density peak is comprised of different particles at different times.

\subsection{External perturber?}

\begin{figure}
\centering
\includegraphics[width=\columnwidth]{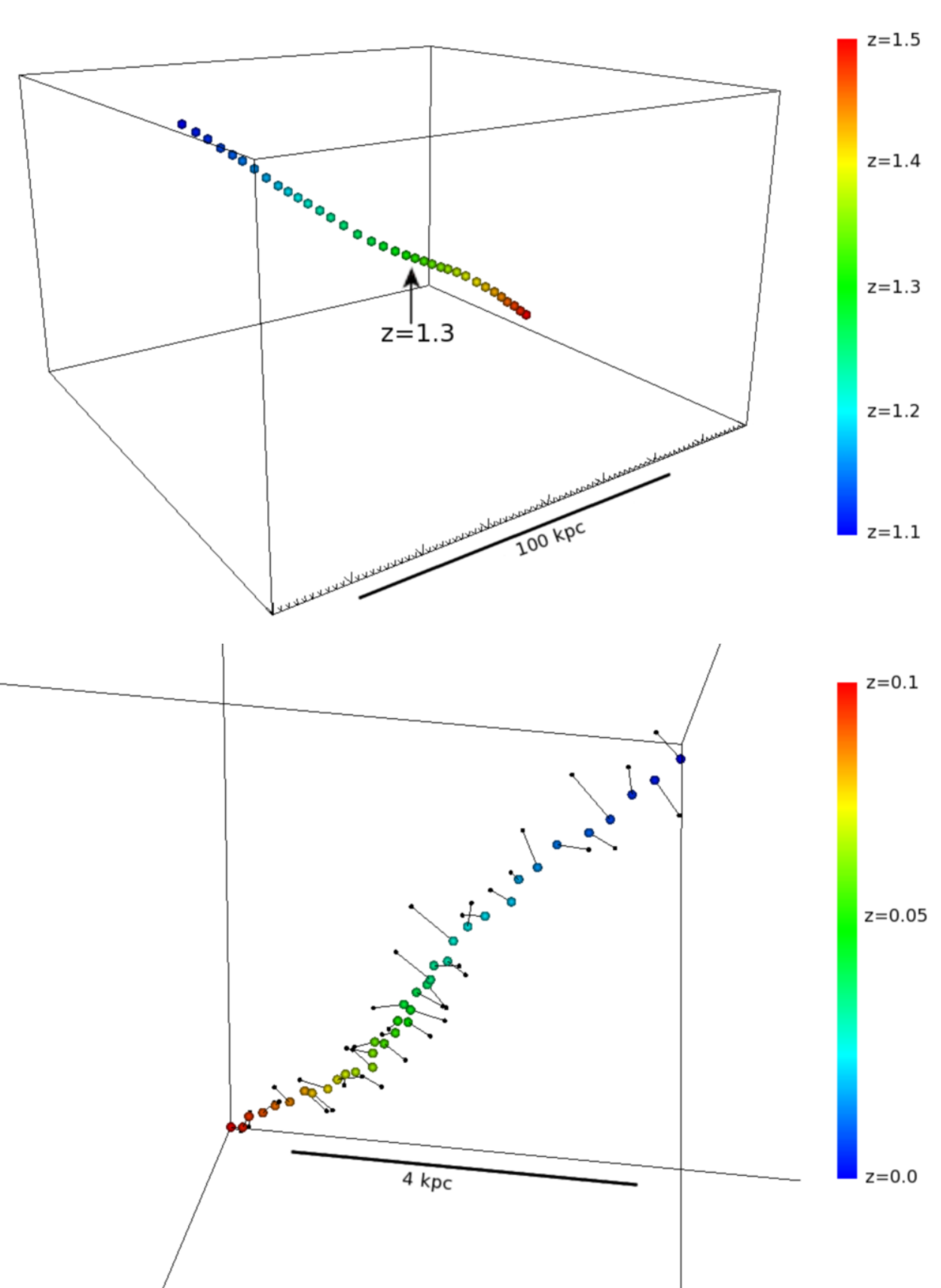}
\caption{Visualizations of the motion of the potential minimum in Eris in absolute (code) units, color coded by redshift. Top: From $z=1.5$ to $z=1.1$. At $z=1.3$ a satellite passes near to the center, but it does not noticeably perturb the dynamical center. Bottom: the last 40 outputs, from $z=0.1$ down to $z=0$. The location of the maximum DM density is marked with a small black sphere connected with a thin line to the position of the potential minimum at that time.}
\label{fig:potmin_motion}
\end{figure}

Next we consider an external perturber as a possible mechanism to give rise to the DM offset. The stellar disk in the center is self-bound, in the sense that it does not require the DM halo to be held together gravitationally. An external perturber, for example in the form of a passing satellite, could then impart a kick to the stellar disk that might cause it to slosh around in the underlying stationary DM halo. In this situation the DM density peak would correspond to the cuspy center of the DM halo, and the offset would simply reflect the displacement of the potential minimum (the stars). One might expect dynamical friction to quickly damp out any such sloshing, but the efficiency of dynamical friction is strongly reduced in a constant density (harmonic) core \citep{read_dynamical_2006}, and thus the center of the self-bound stellar disk may be able to ``orbit'' around the DM density maximum for many dynamical times.

Some fraction of the DM (the low velocity tail) would also be bound to the disk and would move with the stars. In this picture, the existence of a DM offset may be very sensitive to the stellar mass of the galaxy. For a DM halo of a given mass, a higher stellar mass galaxy would bind more of the DM to it, and reduce the strength of the DM offset or eliminated it altogether. Hydrodynamic galaxy formation simulations suffering from a strong baryonic overcooling problem may thus not be able to observe DM offsets.

In Eris we have indeed identified a satellite passing close to the center ($D \approx 65$ kpc) around $z=1.3$, which may be a good candidate for an external perturber, since the DM offset begins to grow around that time. The satellite's mass is $1.8 \times 10^{10} \msun$ at infall ($z=2.7$) and $2.8 \times 10^9 \msun$ at the time of close passage. We have examined the motion of the potential minimum in absolute code coordinates (top panel of Fig.~\ref{fig:potmin_motion}), and do not see any evidence for a sharp kink or abrupt displacement at the time of the satellite's passage. Furthermore, it appears that the potential minimum is moving smoothly in code coordinates, and it is the DM offset that moves around the potential minimum, not the other way around (bottom panel of Fig.~\ref{fig:potmin_motion}). 

\subsection{Density wave excitation by the stellar bar?}

\begin{figure}
\centering
\includegraphics[width=\columnwidth]{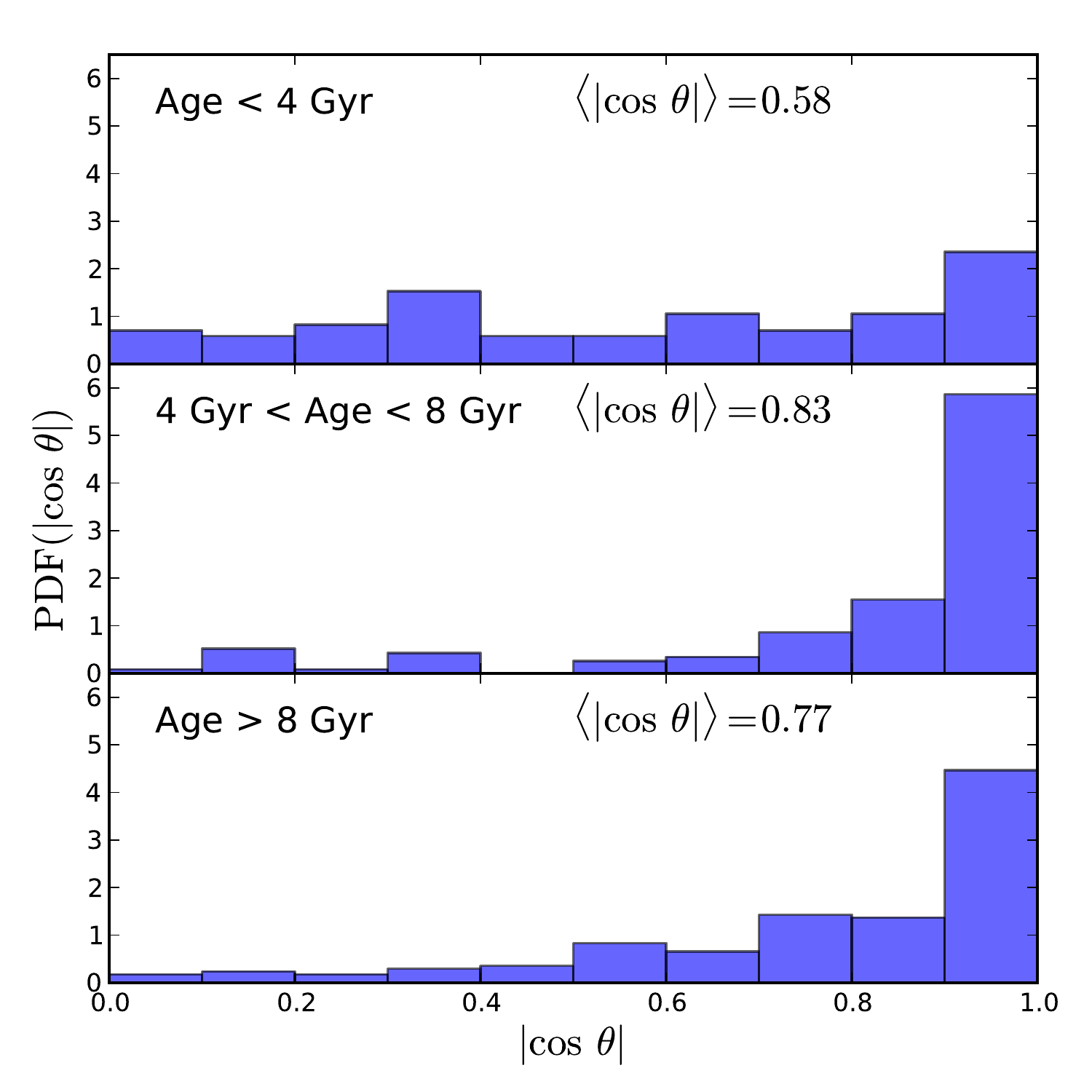}
\caption{Probability distribution functions of the cosine of the angle between the direction towards the DM offset and the orientation of the stellar bar, defined as the major axis of the stellar density ellipsoid inwards of 1 kpc. The three panels show the distributions split by cosmic time: early times, prior to the appearance of the DM offset (\textit{top panel}), an intermediate time period over which the DM offset grows in magnitude (\textit{middle}), and late times, during which the DM offset remains more or less constant at $\sim 340$ pc (\textit{bottom}). The early distribution is consistent with uniform in \costheta, whereas once the DM offset becomes pronounced the distributions prefer large values of \costheta, indicating alignment between the offset peak and the stellar bar.}
\label{fig:bar_alignment}
\end{figure}

Another possibility is that the DM offset is the result of a density wave excitation of some kind. In this explanation the peak would be contributed by different particles at different times, just as observed, its spatial evolution would be set by whatever external source is providing the excitation, and it may be long lived, provided a continuous excitation mechanism exists. A natural candidate for such an external source would be the stellar bar in Eris. The potential is completely dominated by the stars in the central region, and so it seems plausible that the DM distribution may be affected by departures from axisymmetry in the stellar component. Indeed, resonant interactions between the DM halo and a stellar bar have been invoked to explain the transformation of an initially cuspy DM density profile into a cored one \citep{weinberg_bar-driven_2002,weinberg_bar-halo_2007}. In related work, \citet{mcmillan_halo_2005} identified an instability in which a rotating stellar bar pinned to the origin causes the DM cusp to move away from the origin. While they used this effect to argue that the observed flattening of the DM density profile may be artificial, their work provides a proof-of-concept for an off-center DM density peak.

In isolated galaxy simulations it has been possible to identify DM halo particles trapped in different resonances \citep{athanassoula_bar-halo_2002,ceverino_resonances_2007,dubinski_anatomy_2009}. The Inner Lindblad Resonance may lead to the formation of a ``dark bar'' \citep{colin_bars_2006,ceverino_resonances_2007}, which may be showing up in Eris as a DM offset. The ring-like morphology of particles trapped in the co-rotation resonance \citep[see e.g. Fig.11 of][]{ceverino_resonances_2007} is reminiscent of the distribution of DM in the plane of Eris (see Fig.~\ref{fig:slice_all}). A detailed investigation of the resonant structure of DM particle orbits in Eris is beyond the scope of this paper, but will be pursued in future work.

\begin{figure}
\centering
\includegraphics[width=\columnwidth]{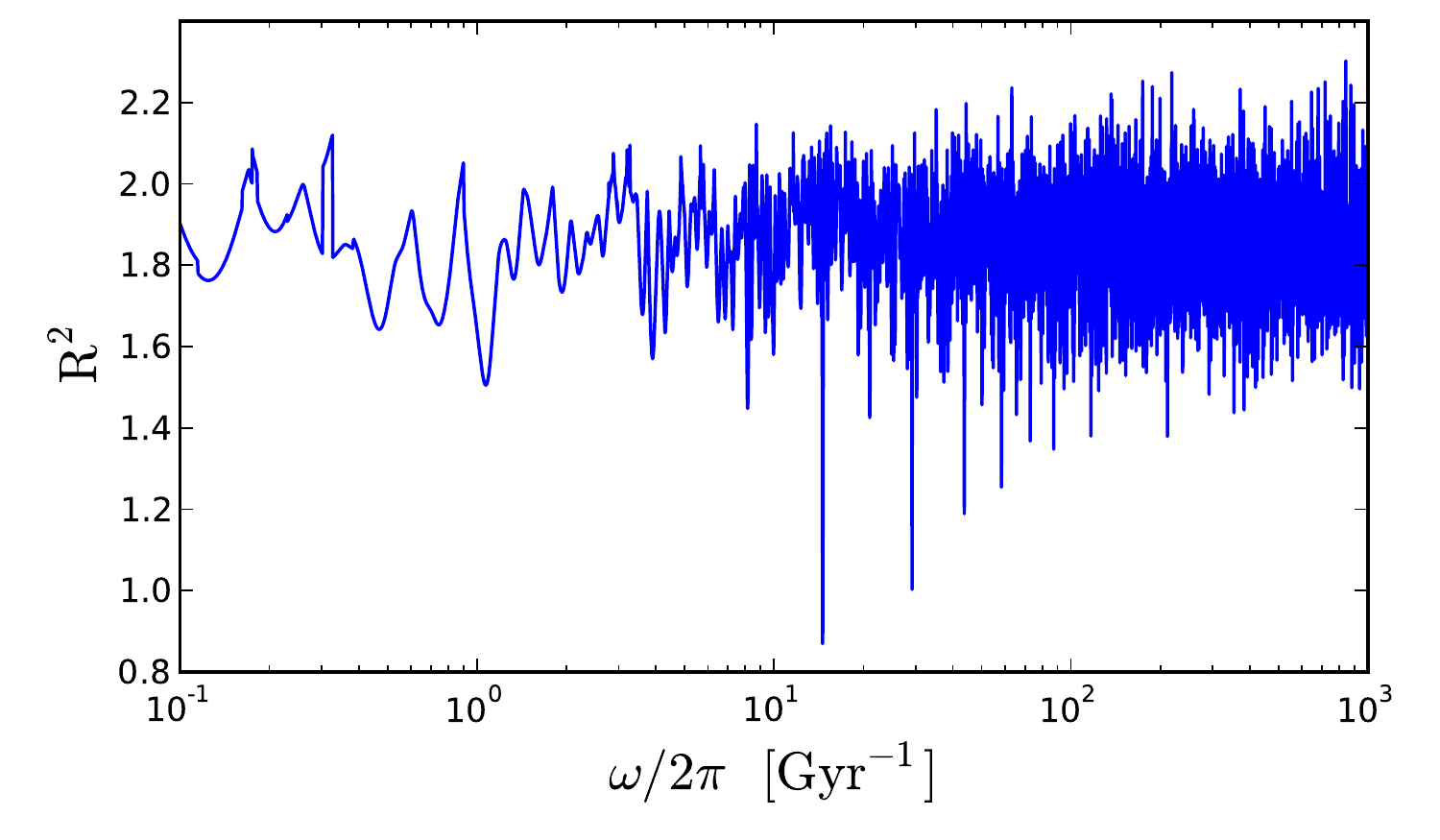}
\includegraphics[width=\columnwidth]{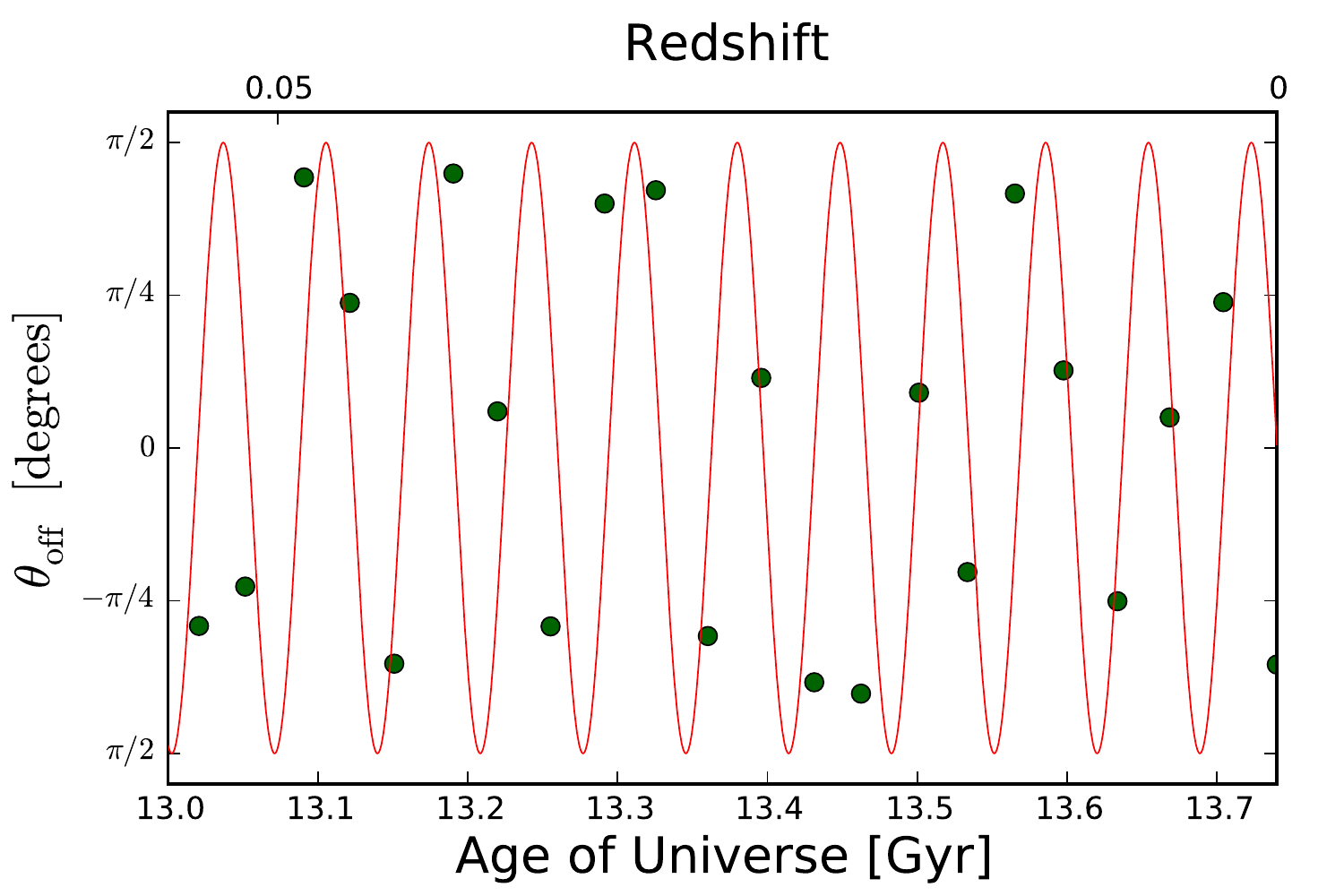}
\caption{Top: Frequency scan of the angular evolution of the direction towards the DM offset. $R^2$ is the sum of the squared differences between the best-fit sinusoid of frequency $\omega$ and the angular position of the DM offset. Lower values of $R^2$ indicate a better fit, and the deep spike to $R^2 < 1$ at $\sim 15 \, {\rm Gyr}^{-1}$ corresponds to a periodicity of 69 Myr in the angular evolution. Bottom: The best-fitting sinusoid (red) overplotted on angular position of the DM offset in the last 22 outputs.}
\label{fig:chi2}
\end{figure}

Arguments in favor of the bar-driven explanation are that the growth of the offset coincides with the formation of a DM core, that the offset peak preferentially lies close to the stellar disk plane, and that the direction from the dynamical center towards the offset density peak appears to be correlated with the orientation of the stellar bar. This last point is demonstrated in Fig.~\ref{fig:bar_alignment}, which shows distributions of \costheta, the angle between the offset peak and the major axis of the stellar density ellipsoid inwards of 1 kpc (i.e. the orientation of the stellar bar).

We have split the distributions up by time epoch: during early times ($< 4$ Gyr), prior to the appearance of a significant DM density offset, the distribution is consistent with uniform in \costheta, corresponding to a completely random placement of the density peak. Between 4 and 8 Gyr the DM offset first begins to become significantly larger than \softening, and the \costheta\ PDF is strongly peaked towards unity. Its mean value of $\langle \costheta \rangle = 0.83$ corresponds to an angle of $34^\circ$. At later times, once the DM offset has become established, the correlation weakens slightly, with $\langle \costheta \rangle$ dropping to 0.77, but it still remains preferentially aligned with the bar.

The top panel of Fig.~\ref{fig:chi2} shows the results of a frequency scan from $\omega/2\pi = 0.1$ to $10^3 \, {\rm Gyr}^{-1}$ of the angular evolution of the direction towards the offset. For every frequency $\omega$, we performed a least squares fit to a sinusoid for all coarsely sampled outputs after 8 Gyr. The plot of $R^2 = \sum_i [\theta_{\rm off}(t_i) - \pi/2 \cos(\omega (t_i - t_o)]^2$ exhibits a distinct low $R^2$ spike at a frequency of $\omega/2\pi = 15 \, {\rm Gyr}^{-1}$, corresponding to a period of 69 Myr, as well as at several of its harmonics ($30, 45, 60 \, {\rm Gyr}^{-1}$, etc.). These spikes indicate the presence of a periodic signal in the offset angle evolution, which is consistent with the alignment of the offset with the orientation of the stellar bar. In the lower panel of Fig.~\ref{fig:chi2} we show the best-fit sinusoid overplotted on the offset angle evolution for the 22 outputs since 13 Gyr.

For completeness we should also mention two problematic aspects of the stellar bar-driven explanation. One is that the strength of stellar bars typically grows and fades with time \citep{bournaud_gas_2002}, modulated by gas accretion. Indeed this is the case in Eris as well: the amplitude of the $m=2$ mode in its stellar disk is strongest at $z>3$, then declines somewhat only to grow again around $z \approx 2$, after which it gradually decreases in strength towards $z=0$ \citep[see Fig.~4 of][]{guedes_pseudobulge_2012}. It is not clear then why the DM offset would saturate at around $300 - 400$ pc in the last 8 Gyr, when the strength of the bar is gradually decreasing. The $\sim 340$ pc extent of the DM offset is also surprisingly small, since Eris' stellar bar is several kpc in length, and its co-rotation radius occurs at 2.7 kpc. It is clear that more work is needed to fully elucidate the role of a stellar bar interaction in explaining the DM offset.

\subsection{Numerical Resolution}

At $\Doff \approx 3 \softening$, the scale of the offset is worrisomely close to the force resolution of the simulation. A detailed study of the internal density structure of the offset peak must certainly await much higher resolution hydrodynamic simulations. Nevertheless, the fact that no offset is observed in ErisDark, which has the same force resolution as Eris, gives us some confidence that the existence of an offset DM density peak is no numerical artifact, and indicates that baryonic physics appears to be responsible in some way.

We would like to be able to check whether the offset remains at the same physical scale ($\sim 340$ pc) even in a simulation with a smaller gravitational softening length. At the moment we do not have access to such a simulation\footnote{We are in the process of re-running the Eris simulation with \softening=50 pc, but this is an expensive simulation that will not finish for several months.}, but we can go in the opposite direction and look in a lower resolution run. We have run ErisLores and ErisDarkLores at eight times poorer mass resolution ($7.8 \times 10^5 \msun$ and $9.5 \times 10^5 \msun$, respectively) and with a four times larger gravitational softening length, \softening=495 pc. We have analyzed these simulations in the same way as described above, except that all length scales have been increased by a factor of four to account for the larger \softening. We CIC-deposit particles onto a grid with cell width of 40 pc and smooth the density field with a Gaussian kernel with $\sigma = \softening = 495$ pc.

As with the higher resolution simulations, we find that the point of maximum density is displaced from the minimum of the potential in ErisLores: $\Doff=430$ pc at $z=0.15$ and 650 pc at $z=0$. And again the DM-only counterpart ErisDarkLores has a much smaller offset, \Doff=60 pc. Although \Doff\ is somewhat larger in ErisLores than in Eris in absolute terms, it is clear that the offset is not scaling linearly with \softening: while $\Doff \approx 3 \softening$ in Eris, it is only 1 - 1.5 \softening\ in ErisLores.

We have also checked time steps and energy conservation in our simulations, as insufficiently short time steps can lead to the formation of a spurious density core \citep{zemp_optimum_2007,dubinski_anatomy_2009}. The Eris simulation suite uses adaptive time steps that scale as $\sqrt{\softening/a}$, where $a$ is the acceleration of a particle, as recommended by \citet{power_inner_2003}. This results in time steps as small as 0.16 Myr at low redshift, which is three times smaller than the fixed time step that \citet{dubinski_anatomy_2009} employed in a simulation with comparable \softening\ (their model m100K) and which they found to be short enough to avoid the formation of an artificial core.

These checks give us further confidence that the DM offset we have observed in Eris cannot be attributed solely to insufficient numerical resolution or timestepping. On the other hand, we cannot claim to have fully resolved \Doff\ either -- higher resolution studies are needed.

\subsection{The role of the central DM core}

It is clear that the central DM core plays an important role for the formation of an off-center DM density peak. Without a flattened central density profile it may be difficult to excite an offset peak, since any non-central density enhancement would be overwhelmed by the steeply rising central cusp. Indeed, as we have seen above, the growth of the DM offset temporally coincides with the formation of such a core. We have not yet determined what physical processes led to the formation of this core in Eris. 

A core formation mechanism that has recently been suggested is the repeated removal of large amounts of gas from the central regions through the action of violent supernovae explosions \citep{read_mass_2005,pontzen_how_2012}. If these baryonic outflows abruptly alter the potential, the DM may respond by flowing out of the center, transforming the cusp into a core in the process. Perhaps this DM redistribution could also create an offcenter DM density peak. Note, however, that the central potential in the Eris galaxy is already dominated by its stellar component by the time the core formation begins at $z \sim 2$. Even though Eris does experience a star burst triggered by a merger at $z \simeq 1$, the associated supernovae feedback is therefore unlikely to substantially alter the central potential. Furthermore, this is a single star formation event, while the mechanism of \citet{pontzen_how_2012} requires multiple application of violent outflows followed by gradual re-accretion of gas.

To further check the role of SN feedback on the formation of a DM density core and peak offset, we have analyzed the last output ($z=0.7$) of the ErisLT run, which is identical to Eris except that it uses a lower star formation threshold (0.1 cm$^{-3}$). This lower SF threshold reduces the efficiency of the supernova-driven gas blowout and results in a disk galaxy with a larger bulge-to-disk ratio and a more compact stellar configuration \citep[see][]{guedes_forming_2011}. The DM distribution, however, appears not to have been affected as dramatically. The DM density profile in ErisLT is almost identical to the one in Eris at $z=0.7$, both exhibiting a core with radius $\sim 1$ kpc, with \Doff=350 pc in ErisLT and \Doff=360 pc in Eris. 

The absence of a dependence in the DM distribution on the star formation threshold suggests that SN-driven outflows are not the only core formation mechanism, highlighting a possible important difference between the evolution of Milky-Way sized galaxies and that of the dwarf galaxies described by \citet{governato_bulgeless_2010,governato_cuspy_2012}. Instead, perhaps the stellar bar may be responsible for this transformation, through resonant angular momentum exchange between the bar and the DM as suggested by \citet{weinberg_bar-driven_2002,weinberg_bar-halo_2007}. Clearly, a more detailed investigation of the DM core formation process in Eris is needed, and will be the topic of a future study.

\section{Discussion}

Provided the central DM distribution in the Eris simulation is representative of that in our Milky Way, the results we have presented here have several important consequences for indirect DM detection efforts aimed at the GC. 

The first concerns the shape of the DM density profile used to predict the annihilation luminosity from the central regions of the Galaxy. Outside of $\sim 1$ kpc the DM density profile in both Eris and ErisDark (as well as VL2 and GHalo) is well described by an NFW or Einasto profile. However, cooling and condensation of gas in Eris has dragged DM towards the center, such that the normalization of the density profile in Eris is larger than in ErisDark: at 1 kpc the mean enclosed DM density is 2.7 times higher in Eris ($0.52 \msun \, {\rm pc}^{-3}$) than in ErisDark ($0.20 \msun \, {\rm pc}^{-3}$). At angles greater than $\sim 7$ degrees from the GC, one may thus expect baryonic processes to lead to a significant enhancement of the surface brightness of diffuse radiation arising from DM annihilation.

As discussed in Sec.~\ref{sec:DM_core}, at even smaller radii baryonic physics leads to the formation of a DM core in Eris, and as a result the mean enclosed DM density in Eris never rises to more than $0.73 \msun \, {\rm pc}^{-3}$. For comparison, an extrapolation of the best-fitting Einasto profile for ErisDark reaches 1 $\msun \, {\rm pc}^{-3}$ at 200 pc, 10 $\msun \, {\rm pc}^{-3}$ at 3 pc, and asymptotes to 90 $\msun \, {\rm pc}^{-3}$. We thus caution against extrapolating density profiles determined from dissipationless DM-only simulations (such as Via Lactea II, GHalo, and Aquarius) all the way in to the GC in order to infer central annihilation luminosities.

\begin{figure}
\centering
\includegraphics[width=\columnwidth]{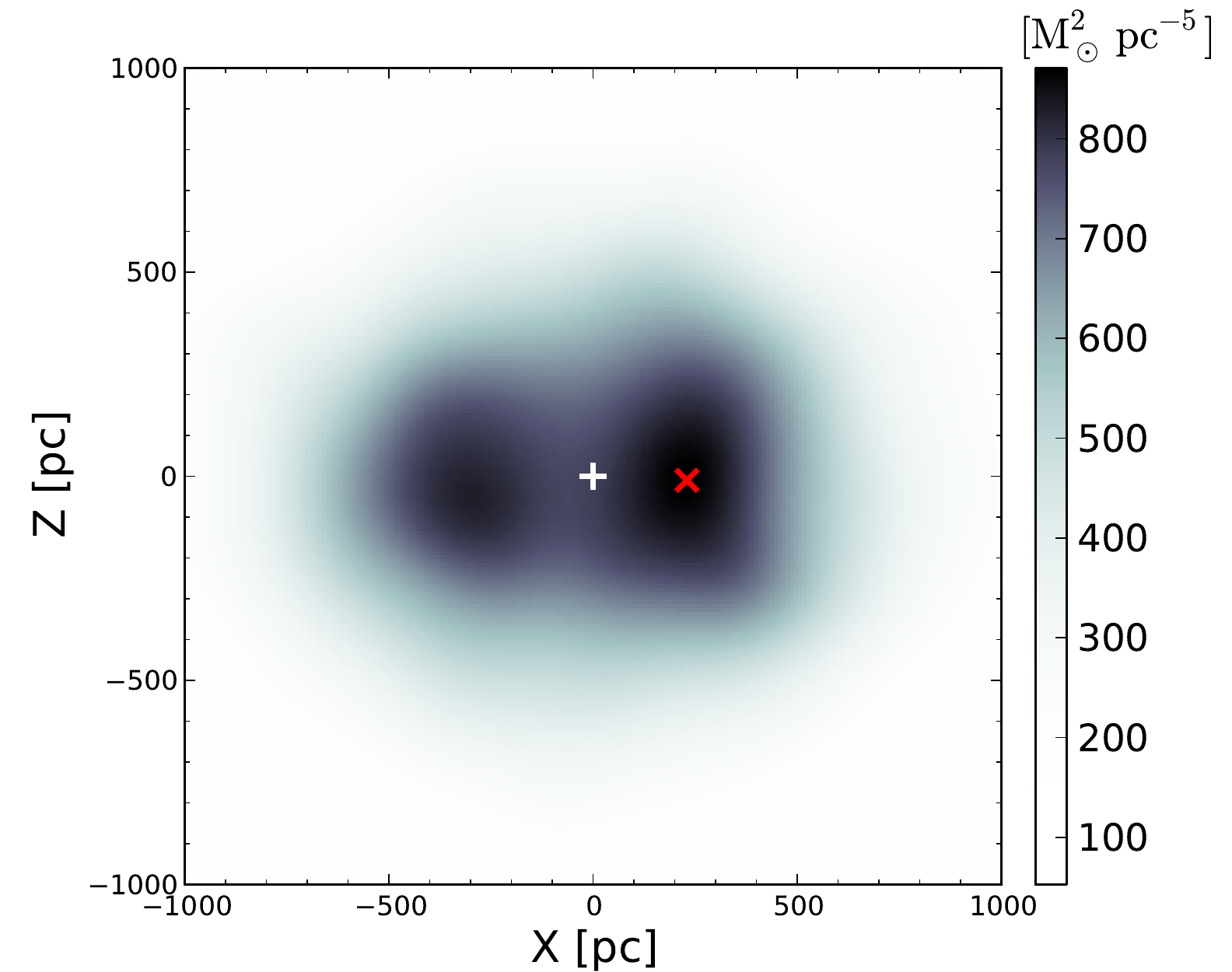}
\caption{DM annihilation luminosity ``surface density'' ($\int \! \rho_{\rm DM}^2 \, {\rm d}\ell$) in the central $2 \, {\rm kpc} \times 2 \, {\rm kpc}$ region of Eris. The contrast between the dynamical center (white plus) and the peak (red cross) is $\sim 11\%$. The vertical axis corresponds to the disk normal, and the coordinate system has been rotated to maximize the angular offset of the peak.}
\label{fig:DMannihilation}
\end{figure}

Secondly, a maximum in the DM density that is not coincident with the dynamical center of the Galaxy significantly modifies expectations for where the central surface brightness of DM annihilation radiation should peak. The time-averaged value of \Doff\ in Eris is 340 pc, which, if seen edge on, would correspond to an angular offset of 2.4 degrees. For comparison, the search for a DM annihilation signal by the H.E.S.S. imaging Atmospheric Cherenkov Telescope was restricted to a region of $45 - 150$ pc in projected distance from the GC \citep{abramowski_search_2011}. In this context, the recent report of a highly significant detection of a gamma-ray line at $127 \pm 2$ GeV in Fermi data from the GC \citep{su_strong_2012} is particularly intriguing. \citet{su_strong_2012} report that the signal is maximized at a Galactic longitude of $\ell \approx 1.5$ degrees. While this offset may have initially been viewed as a strike against a DM annihilation interpretation of the line, our work demonstrates as a proof-of-principle that just such an offset should perhaps be expected.

In Fig.~\ref{fig:DMannihilation} we show the DM annihilation luminosity ``surface density'',  $\int \! \rho_{\rm DM}^2  {\rm d}\ell$, calculated from the central $(2 \, {\rm kpc})^3$ region in Eris. The view is through the disk plane (the vertical axis corresponds to the disk normal) and the coordinate system has been rotated such that the offset peak is viewed edge-on, i.e. maximizing the angular offset. As before, the 3D density grid was first smoothed with a Gaussian kernel of width $\sigma = \softening$. We then squared it, multiplied the cells by their volume to get a luminosity, summed them up along the Y-axis, and finally divided the resulting map by the area of each cell to get a surface density\footnote{To further clarify, the sum of all $200^2$ ``pixels'' multiplied by the total area ($4 \times 10^6 \, {\rm pc}^2$) is equal to the total luminosity emitted from the $(2 \, {\rm kpc})^3$ cube.} in units of ${\rm M}_\odot^2 \, {\rm pc}^{-5}$. This map does not take into account any luminosity produced by DM along the line of sight outside of the central cube. 

The contrast between the maximum of the map ($870 \msun^2 \, {\rm pc}^{-5}$) and the center ($780 \msun^2 \, {\rm pc}^{-5}$) is only about 11\% at $z=0$. Furthermore there is a second sub-dominant peak to the left of the center with a luminosity surface density of $830 \msun^2 \, {\rm pc}^{-5}$, only 5\% less than the global maximum. If anything the contrast at $z=0$ is a bit of an outlier, as the mean contrast between offset peak and GC over the past 4 Gyr is only $\sim 5\%$ with a 3\% standard deviation. The greatest contrast measured over all outputs is 15\%. Such small contrasts may not be compatible with the measurement of \citet{su_strong_2012}, but it is important to keep in mind that the internal structure of the peak is certainly not resolved in the simulation. It is conceivable that the contrast would increase with higher resolution, especially if resonances with the stellar bar are responsible, since those would be spread out artificially at low resolution.

\section{Conclusions}

We have analyzed the distribution of DM in the central regions of the hydrodynamical galaxy formation simulation Eris, one of the highest resolution and most realistic simulations to date of the formation of a barred spiral galaxy like our own Milky Way. Surprisingly, we find that the peak of the DM density in Eris is typically offset from its dynamical center by several hundred parsec. No such offset is observed in its DM-only twin simulation ErisDark, nor in the much higher resolution DM-only Via Lactea II and GHalo simulations.

The DM offset in Eris begins to appear around $z=1.5$ and grows over a period of 2 Gyr to a stable value of $\langle \Doff \rangle = 340$ pc (almost three gravitational softening lengths), with a dispersion of 50 pc. The onset and duration of the DM offset appears to be well correlated with the formation of a nearly constant DM density core. The distributions of $\rho_{\rm max}$ and $\Doff$ over the past 4 Gyr are inconsistent with a statistical fluctuation. Neither is the density peak a gravitationally bound structure, which rules out an incompletely disrupted subhalo core as an explanation. The most likely explanation may be a density-wave-like excitation by the stellar bar, possibly related to the resonant mechanism proposed by \citet{weinberg_bar-driven_2002, weinberg_bar-halo_2007} to explain the transformation of a central DM cusp into a core. Arguments in favor of this explanation are the fact that the DM offset appears preferentially near the disk plane, that it is aligned to $\sim 30$ degrees with the orientation of the stellar bar, and that is shows a periodicity of $\sim 70$ Myr.

A central DM offset is of particularly interest in the context of the recent report by \citet{su_strong_2012} of a highly significant detection of gamma-ray line emission from a region $\sim 1.5$ degrees ($\sim 200$ pc projected) away from the Galactic Center. At first impression such a large angular offset would seem to argue against a DM annihilation interpretation of this signal. Our work demonstrates that in fact just such an offset should perhaps be expected. We note, however, that at the current resolution of our numerical simulations the low contrast ($5 - 15\%$) of the annihilation surface brightness between the offset peak and the GC may be too small to accommodate a DM annihilation explanation.

We conclude by acknowledging that properly resolving the effects of baryonic physics on the central DM distribution, in particular those involving resonant bar-halo interactions, requires much higher resolution than we have been able to afford so far in cosmological hydrodynamics simulations. Further studies at higher resolution and exploring different baryonic physics implementation are sorely needed. Of particular importance are clarifying the role of the star formation threshold parameter and supernovae feedback for the formation of a DM core, as well as the influence of the supermassive black hole at the GC on the DM distribution.

\acknowledgments
MK thanks Doug Finkbeiner for initially suggesting this analysis and several encouraging exchanges afterwards, and Andrey Kravtsov, Daniel Ceverino, Anatoly Klypin, Eliot Quataert, and Justin Read for valuable discussion of these results. JG was partially funded by the ETH Zurich Postdoctoral Fellowship and the Marie Curie Actions for People COFUND Program. The high time resolution re-reun of Eris was performed on the Cray XE6 Monte Rosa at the Swiss National Supercomputing Centre (CSCS). This work was supported in part by a grant from the Swiss National Supercomputing Centre (CSCS) under project IDs s205 and s352, and in part by the U.S. National Science Foundation, grants OIA-1124453 (PI P.~Madau) and OIA-1124403 (PI A.~Szalay).

\bibliographystyle{apj}
\bibliography{DM_Offset}

\end{document}